\documentclass[12pt]{article}
\pdfoutput=1

\usepackage[bulletsep]{collref}
\usepackage{amssymb,graphicx}
\usepackage[intlimits]{amsmath}
\usepackage{bbm}
\usepackage[small]{subfigure}

\usepackage{MnSymbol}

\makeatletter \@addtoreset{equation}{section} \makeatother

\makeatletter
\let\old@startsection=\@startsection
\let\oldl@section=\l@section
\renewcommand{\@startsection}[6]{\old@startsection{#1}{#2}{#3}{#4}{#5}{#6\mathversion{bold}}}
\renewcommand{\l@section}[2]{\oldl@section{\mathversion{bold}#1}{#2}}
\makeatother

\makeatletter
\let\old@makecaption=\@makecaption
\def\@makecaption{\small\old@makecaption}
\makeatother

\usepackage{mathtools}
\DeclarePairedDelimiter\galscal{\langle}{\rangle_w}

\renewcommand{\geq}{\geqslant}

\newcommand{\av}[1]{\left\llangle #1 \right\rrangle}
\newcommand{\eqq}{\stackrel{!}{=}}

\usepackage[hidelinks]{hyperref}
\begin{document}

\newcommand{\be}{\begin{equation}}\newcommand{\ee}{\end{equation}}
\newcommand{\bea}{\begin{eqnarray}} \newcommand{\eea}{\end{eqnarray}}
\def\p{\partial}
\def\pa{\partial}
\def\ov{\over }
\def\a{\alpha }
\def\g{\gamma}
\def\s{\sigma }
\def\td{\tilde }
\def\vp{\varphi}
\def\gd{\nu }
\def \ha {{1 \over 2}}

\def\KK{{\cal K}}

\def\Xint#1{\mathchoice
{\XXint\displaystyle\textstyle{#1}}
{\XXint\textstyle\scriptstyle{#1}}
{\XXint\scriptstyle\scriptscriptstyle{#1}}
{\XXint\scriptscriptstyle\scriptscriptstyle{#1}}
\!\int}
\def\XXint#1#2#3{{\setbox0=\hbox{$#1{#2#3}{\int}$ }
\vcenter{\hbox{$#2#3$ }}\kern-.5\wd0}}
\def\ddashint{\Xint=}
\def\dashint{\Xint-}

\newcommand\cev[1]{\overleftarrow{#1}}

\begin{flushright}\footnotesize
\texttt{NORDITA-2019-003} \\
\texttt{UUITP-1/19}\\
\texttt{ICCUB-19-001}

\vspace{0.6cm}
\end{flushright}

\renewcommand{\thefootnote}{\fnsymbol{footnote}}
\setcounter{footnote}{1}

\begin{center}
{\Large\textbf{\mathversion{bold}   $N=2^*$ Phase Transitions and Holography }
\par}

\vspace{0.8cm}

\textrm{Jorge~G.~Russo$^{1,2}$, Erik~Wid\'en$^{3,4}$ and
Konstantin~Zarembo$^{3,4}$\footnote{Also at ITEP, Moscow, Russia}}
\vspace{4mm}

\textit{${}^1$ Instituci\'o Catalana de Recerca i Estudis Avan\c cats (ICREA), \\
Pg. Lluis Companys, 23, 08010 Barcelona, Spain.}\\
\textit{${}^2$ Departament de F\' \i sica Cu\' antica i Astrof\'\i sica and ICCUB\\
Universitat de Barcelona, Mart\'i Franqu\`es, 1, 08028
Barcelona, Spain. }\\
\medskip
\textit{${}^3$Nordita, KTH Royal Institute of Technology and Stockholm University,
Roslagstullsbacken 23, SE-106 91 Stockholm, Sweden.}\\
\textit{${}^4$Department of Physics and Astronomy, Uppsala University\\
SE-751 08 Uppsala, Sweden.}\\
\vspace{0.2cm}
\texttt{jorge.russo@icrea.cat, erikwid@kth.se, zarembo@nordita.org}

\vspace{3mm}


\par\vspace{1cm}

\textbf{Abstract} \vspace{3mm}

\begin{minipage}{13cm}

We clarify the relationship between probe analysis of the supergravity dual and the large-$N$ solution of the localization matrix model for the planar $\mathcal{N}=2^{*}$ super-Yang-Mills theory. A formalism inspired by supergravity allows us to systematically solve the matrix model at strong coupling. Quite surprisingly, we find that quantum phase transitions, known to occur in the $\mathcal{N}=2^{*}$ theory, start to be visible at the third order of the strong-coupling expansion and thus constitute a perturbative phenomenon on the string worldsheet.

\end{minipage}

\end{center}

\vspace{0.5cm}



\setcounter{page}{1}
\renewcommand{\thefootnote}{\arabic{footnote}}
\setcounter{footnote}{0}

\section{Introduction}

The ${\mathcal N}=2^*$ theory is obtained by softly breaking the ${\mathcal N}=4$ super Yang-Mills theory to ${\mathcal N}=2$ supersymmetry 
by the addition of a mass term. It provides an extremely interesting setup to understand holography in non-conformal systems, as the
precise supergravity dual is known \cite{Pilch:2000ue,Buchel:2000cn}. 
 One well known feature of  the  ${\mathcal N}=4$ theory is its smooth dependence
on the 't~Hooft coupling $\lambda=g_{{\rm YM}}^{2}N $ all the way from $\lambda=0 $ to $\lambda =\infty$. This feature changes dramatically when the theory is deformed. The resulting ${\mathcal N}=2^*$ theory undergoes an infinite number
of phase transitions \cite{Russo:2013qaa,Russo:2013kea}. The first transition occurs at $\lambda \approx 35. 42$ and is followed by a sequence of secondary transitions appearing at regular intervals   $\sqrt{\lambda}\approx \pi n$ with integer $n\gg 1$. The critical behavior is caused by rearrangement of the vacuum  whereby new massless resonances enter the spectrum \cite{Russo:2013qaa}. 

Holography allows one to explore the vacuum structure at strong coupling by placing a D3-brane probe in the dual supergravity background \cite{Buchel:2000cn,Evans:2000ct}.  Quite remarkably, the eigenvalue distribution predicted by the supergravity analysis obeys Wigner semicircle law \cite{Buchel:2000cn}, a hallmark of random matrix theory. This finding led the authors of \cite{Carlisle:2003nd} to speculate that Gaussian random matrices underlie vacuum structure of the $\mathcal{N}=2^{*}$ SYM and other strongly-coupled $\mathcal{N}=2$ gauge theories at large-$N$. We now know that random matrices arise in $\mathcal{N}=2$ theories upon supersymmetric localization, which maps the partition function on $S^{4}$ to a zero-dimensional matrix model \cite{Pestun:2007rz}. Albeit the matrix model is not Gaussian in the $\mathcal{N}=2^{*}$  case (it is only such for $\mathcal{N}=4$ SYM \cite{Erickson:2000af,Drukker:2000rr,Pestun:2007rz}), its  course-grained, averaged eigenvalue distribution indeed takes semicircular Wigner shape at strong coupling, in spectacular confirmation of the holographic predictions \cite{Buchel:2013id}.

The gauge/gravity duality for the ${\mathcal N}=2^*$ theory has withstood many other remarkable, rigorous tests, 
by confronting the exact results obtained from supersymmetric localization with the corresponding  gravity dual observables, such as Wilson loops \cite{Buchel:2013id,Chen-Lin:2015xlh,Chen-Lin:2017pay,Bobev:2018hbq} and free energy on $S^{4}$ \cite{Bobev:2013cja}. 
Matrix model tools \cite{Buchel:2013id,Russo:2013qaa,Russo:2013kea,Russo:2013sba,Chen:2014vka,Zarembo:2014ooa,Chen-Lin:2015dfa}, advances in supergravity \cite{Balasubramanian:2013esa,Bobev:2013cja,Bobev:2018hbq} and deeper understanding of string \cite{Dimov:2003bh,Buchel:2013id,Chen-Lin:2017pay} and  D-brane \cite{Buchel:2000cn,Evans:2000ct,Chen-Lin:2015xlh} probes   have been instrumental in these developments. While localization presents abundant evidence for phase transitions \cite{Russo:2013qaa,Russo:2013kea,Chen:2014vka,Zarembo:2014ooa,Chen-Lin:2015dfa,Russo:2014nka,Russo:2015vva}, they so far remained in limbo on the gravity side.

A challenge is to understand  how the non-analytic features of the quantum phase transitions are reflected in the holographic description.
Strictly speaking, the supergravity background describes
the $\lambda=\infty $ theory. In this limit, 
the gauge theory has infinitely many phases and it approaches an average, coarse-grained, smooth description where
the non-analytic behavior does not show up in the leading approximation  \cite{Russo:2013qaa,Russo:2013kea}.
For example, the vacuum expectation value of the 1/2 BPS circular Wilson loop approaches
$\ln\langle W(C)\rangle\approx ML_{C}\sqrt{\lambda }$, which would seem to have a continuous dependence on the coupling. This 
average description of the strong coupling regime makes particularly difficult and challenging to reveal
the existence of phase transitions  on the string theory side of the duality.

Keeping this mind, we revisit the supergravity analysis of the vacuum structure \cite{Buchel:2000cn} in a hope to establish a tighter connection to localization. A closer look at the supergravity calculation reveals striking similarities to the strong-coupling solution of the matrix model \cite{Zarembo:2014ooa}. Armed with this observation we extend and streamline large-$N$ analysis of the localization matrix model and push the strong-coupling solution to the third order in $1/\sqrt{\lambda }$. Quite surprisingly, phase transitions become fully visible  starting with this order enabling, for example, computation of the critical indices.

The upshot of our analysis is that the phase transitions constitute a perturbative phenomenon from the supergravity/string theory point of view. They were not visible so far for a simple reason that existent string or gravity calculations have explored the LO and NLO of the strong-coupling expansion, while the phase transitions appear at NNLO.

\section{Probe analysis and eigenvalue density}\label{sec:sugrapred}

The probe analysis of the Pilch-Warner geometry defines an  exact distribution of D3-branes on the enhan\c{c}on
\cite{Buchel:2000cn}, which appears to match the
density of eigenvalues in the gauge theory, obtained from localization \cite{Buchel:2013id,Russo:2013qaa,Russo:2013kea}. 
Here we review the supergravity calculation, slightly generalizing it to include finite-mass effects.
This is crucial for potential non-analytic features responsible for the phase transitions. We will later incorporate elements of the supergravity analysis in the solution of the localization matrix model.

In the gauge-theory language, a probe D3-brane placed at position $u$ on the locus of marginal stability corresponds to a point on the moduli space of vacua of the $SU(N+1)$ theory parametrized  by
\be
\Phi 
= {\rm diag}( a_1- u/N, a_2-u/N,...., a_N-u/N, u )\ ,\qquad \sum_i a_i =0,
\ee
where $\left\{a_{i}\right\}$ describe the vacuum of the $SU(N)$ theory without the probe, sourced by the geometry itself. The kinetic term of the probe is characterized by the effective coupling
\begin{equation}\label{tau-gauge}
 \tau =\frac{4\pi i}{g_{{\rm YM}}^{2}}+\frac{\theta }{2\pi }
 =\frac{\partial^2 {\cal F}}{\partial u^2}\,,
\end{equation}
where $\mathcal{F}$ is the prepotential.
The instanton part of the prepotential is irrelevant in the large $N$ limit, so we just need to include the classical and 
one-loop parts. These are given by
\bea\label{prepot}
{\cal F}(u) &=& \frac{2\pi Ni}{\lambda}\, u^2 +\frac{i}{8\pi }\sum_i \left[ 2 (u-a_i)^2 \ln (u-a_i)^2
\right.
\nonumber\\
&& 
\left.-(u-a_i-M)^2 \ln (u-a_i-M)^2- (u-a_i+M)^2 \ln (u-a_i+M)^2\right],
\eea
where we have omitted an additive term that does not depend on $u$ and thus does not contribute to the effective coupling.

Expanding the probe action to quadratic order in the field
strength, one can identify the effective coupling with the supergravity axi-dilaton:
\begin{equation}
 \tau =\,{\rm e}\,^{-\phi }+C_{0},
\end{equation}
It was verified in \cite{Buchel:2000cn} that the locus where the probe experiences no force is real two-dimensional (one complex dimensional). Evaluating the axi-dilaton on the locus of marginal stability, \cite{Buchel:2000cn} found that
\be
\label{sugratau}
\tau_{\rm sugra} = \frac{4\pi Ni}{\lambda } \,\,\frac{u}{\sqrt{u^{2}-\mu ^{2}}}\,.
\ee
On the other hand, from (\ref{tau-gauge}) and (\ref{prepot}):
\be
\label{gaugetau}
\tau = \frac{4\pi  Ni}{\lambda} +\frac{i}{2\pi }\sum_i  \ln \frac{(u-a_i)^2}{|(u-a_i)^2-M^2|}\ .
\ee

The eigenvalue density can be computed by comparing (\ref{sugratau}) and (\ref{gaugetau}).
This calculation was carried in \cite{Buchel:2000cn} using the approximation $M^2\ll (u-a_i)^2$. Such an approximation is justified at strong coupling, because, as we shall see, $a_{i}\sim u\sim \sqrt{\lambda }\,M$.
In this approximation,
\be\label{taugauge1}
\tau \approx \frac{4\pi Ni}{\lambda} +\frac{i}{2\pi }\sum_i   \frac{M^2}{(u-a_i)^2}
=\frac{4\pi Ni}{\lambda} +\frac{M^{2}Ni}{2\pi }\int_{-\mu}^\mu  \frac{dx\,\rho(x)}{(u-x)^2}\, .
\ee
Taking discontinuity across the cut and comparing to (\ref{sugratau}),
one finds 
\be
M^2 \rho'(x) \approx - \frac{8\pi }{\lambda} \frac{x}{\sqrt{\mu^2-x^2}}\,.
\ee
This gives
\be\label{Wigner1}
\rho(x) \approx \frac{8\pi}{\lambda M^2} \sqrt{\mu^2-x^2}\,,
\ee
the result found in \cite{Buchel:2000cn}. It is easy to check that not only the discontinuity matches, but the whole function (\ref{sugratau}) is reproduced by the Wigner density.
The width of the eigenvalue distribution is fixed by normalization:
\be
\mu = \frac{\sqrt{\lambda}M}{2\pi}\, ,
\label{mumu}
\ee
which justifies {\it a posteriori} the simplifying assumption that $M^2\ll (u-a_i)^2$.
It would have been technically cleaner to solve for the density without making this assumption and to expand in $M$ only after establishing that $\mu \gg M$. Let us repeat the calculation without making approximations in intermediate steps.

Differentiating (\ref{gaugetau}) with respect to $u$, we obtain
\be
\label{gaugeprime}
\tau '(u) =  \frac{Ni}{2\pi }\int_{-\mu}^{\mu} dx \ \rho(x) \left( \frac{2}{u-x}- \frac{1}{u-x+M}- \frac{1}{u-x-M}\right).
\ee
This is to be compared with the supergravity prediction, obtained by differentiating  (\ref{sugratau}):
\be
\label{sugraprime}
\tau_{\rm sugra}'(u) =- \frac{4\pi N i}{\lambda }\, \,\frac{\mu ^{2}}{(u^2-\mu^2)^{3/2}}\,.
\ee
Equating, as before, discontinuities of the two functions, we get:
\begin{equation}\label{first-findiff}
  2\rho(x) -\rho(x+M)-\rho(x-M)=\frac{8\pi }{\lambda }\, \,\frac{\mu ^{2}}{(\mu ^2-x^2)^{3/2}}\,.
\end{equation}

Taylor expanding in $M$  approximates the left-hand side
to $- M^2\rho''(x)$, resulting in a differential equation solved by the Wigner distribution (\ref{Wigner1}). The finite-difference equation can be solved as well, without the small-mass approximation. The technical details are postponed till the next section, but the key qualitative features are apparent just from the form of the equation. The boundary conditions on the density are set at $x=\pm \mu $. In between of those points the evolution goes in steps of $M$. It thus crucially depends on how many steps separate the endpoints of the interval. If one gets from $-\mu $ to $\mu $ in an integer number of steps, namely when $2\mu =nM$, a resonance occurs and the solution undergoes a discontinuous mutation. These are precisely the phase transitions visible on the gauge-theory side \cite{Russo:2013qaa}.

Still, it is fair to say that there is no direct evidence of phase transitions from the pure gravitational description. The gravitational input is (\ref{sugraprime}), a nice continuous function of $\mu $. Irregularities appear when the gravitational data is recycled  into the eigenvalue density which, by itself, does not have any direct holographic description.  A clear signature of the phase transitions would be a non-analytic behavior of the expectation values (moments of the eigenvalue density). Such non-analyticities do not occur in the supergravity approximation, a property which we will be able to quantify by a systematic strong-coupling calculation on the gauge-theory side.

The supergravity analysis suggests  that $\tau '(u)$ is a more convenient characterization of the eigenvalue distribution than the density. The latter has a very irregular structure at strong coupling \cite{Chen:2014vka,Zarembo:2014ooa}. An ansatz, first proposed in \cite{Zarembo:2014ooa,Anderson:2014hxa}, expresses it through a regular function bearing a lot of similarity to (\ref{sugraprime}). The ansatz can be shown to  solve the finite-difference equation (\ref{first-findiff}), and
in the next section we sharpen and extend this connection between supergravity and gauge-theory quantities.

\section{Saddle point equation}

The partition function of the $\mathcal{N}=2^{*}$ theory on $S^{4}$ can be computed exactly by supersymmetric localization \cite{Pestun:2007rz}:
\begin{equation}\label{eigenint}
 Z=\int_{}^{}d^{N-1}a\,\prod_{i<j}^{}\frac{(a_{i}-a_{j})^{2}H(a_{i}-a_{j})}
 {H(a_{i}-a_{j}+M)H(a_{i}-a_{j}-M)}\,
 \,{\rm e}\,^{-\frac{8\pi ^{2}N}{\lambda }\sum_{i}^{}a_{i}^{2}}
 \left|\mathcal{Z}_{{\rm inst}}\right|^{2},
\end{equation}
where $a_{i}$ are the eigenvalues of the scalar from the vector multiplet:
\begin{equation}
 \Phi =\mathop{\mathrm{diag}}(a_{1}\ldots a_{N})
\end{equation}
and
\begin{equation}
 H(x)=\prod_{n=1}^{\infty }\left(1+\frac{x^{2}}{n^{2}}\right)^{n}
 \,{\rm e}\,^{-\frac{x^{2}}{n}}
\end{equation}
The instanton contribution can be neglected at large-$N$\footnote{The instanton weight $\,{\rm e}\,^{-8\pi ^{2}N/\lambda }$ is exponentially small at large-$N$, but the moduli integral can potentially overcome the exponential suppression \cite{Gross:1994mr}. It has been explicitly checked that the instanton explosion never happens in the planar $\mathcal{N}=2^{*}$ theory \cite{Russo:2013kea}.}, while the remaining integral is of the saddle-point type and is solved by  a configuration of the eigenvalues that minimizes the effective action.

The eigenvalue integral (\ref{eigenint}) is written in terms of the dimensionless variables. The canonical dimensions are recovered by rescaling $a_{i}\rightarrow a_{i}R$, $M\rightarrow MR$, where $R$ is the radius of the four-sphere. In the decompactification limit $R\rightarrow \infty $ the problem boils down to solving a singular integral equation\cite{Russo:2013kea}\footnote{Which arises upon differentiating the original saddle-point equation twice and using the large-argument asymptotics $H(x)\approx |x|^{-x^{2}}$.}
\begin{equation}\label{intnormal1}
 \strokedint_{-\mu }^\mu
 dy\,\rho (y)  \left( \frac{2}{x-y} - \frac{1}{x-y+M}-\frac{1}{x-y-M} \right) =0,
\end{equation}
supplemented by two normalization conditions:
\begin{eqnarray}\label{formintegrated}
 &&\int_{-\mu }^{\mu }dx\,\rho (x)\ln\frac{\left|M^2-x^2\right|}{x^2}=\frac{8\pi ^2}{\lambda }
 \\
 &&\int_{-\mu }^{\mu }dx\,\rho (x)=1. \label{unitnormdens}
\end{eqnarray}
The eigenvalue density is defined on the interval $(-\mu ,\mu )$ and has the inverse square root singularities at its endpoints. The symmetry-breaking mass scale $\mu $ is a function of the 't~Hooft coupling, implicitly determined by the auxiliary normalization conditions. A known exact solution of these equations in terms of theta-functions \cite{Russo:2013qaa,Russo:2013kea} is valid for sufficiently small  $\lambda$ and terminates at a quantum phase transition at $\lambda _{c}\approx 35.42$. Remarkably, the same  
$\lambda_c$ appears to be a special value of the coupling for all $SU(N)$ theories with  $N\geq 3$, where an exact, analytic solution is also known up
to $\lambda_c$ 
\cite{Hollowood:2015oma}. On the other hand, the theory with $SU(2)$ gauge group has a smooth behavior with $\lambda $ all the way from
0 to infinity \cite{Russo:2014nka,Hollowood:2015oma}.

At large $\lambda $, as discussed above, the eigenvalue density approaches 
\be
\rho(x) \approx \frac{2}{\pi \mu^2} \sqrt{\mu^2-x^2}\ ,
\ee
with
\begin{equation}\label{mulambda}
 \mu =\frac{\sqrt{\lambda }\,M}{2\pi } \qquad \left(\lambda \rightarrow \infty \right).
\end{equation}
which matches the prediction from the Pilch-Warner supergravity solution.
This is, however, a coarse-graining description which hides a complicated structure involving multiple 
discontinuities.

In general, the structure of the density qualitatively changes when the length of the eigenvalue interval crosses an integer multiple of $M$. A new resonance appears at that point and the system undergoes a phase transition. It is convenient to introduce variables $n$ and $\Delta $ that characterize the size of the interval relative to $M$ \cite{Anderson:2014hxa}:
\begin{equation}\label{n-Delta}
 n=\left[\frac{2\mu }{M}\right],\qquad \Delta =\left\{\frac{2\mu }{M}\right\},\qquad 2\mu =(n+\Delta)M ,
\end{equation}
where $[x]$, $\{x\}$ denote integer/fractional part of $x$.  An integer $n$ enumerates phases of the theory.

The resonance terms in the saddle-point equation (\ref{intnormal1}) induce singularities in the eigenvalue distribution, at points $\mu -lM$ and $-\mu +lM$ with $l=1\ldots n$. As a result, the density has a complicated, irregular structure, especially at strong coupling, featuring a growing number of peaks and jumps. But, as we already discussed, the density is not the most convenient characterization of the eigenvalue distribution.

Inspired by the supergravity analysis we introduce a resolvent-type function
\begin{equation}
 r(u)=\frac{i}{2\pi }\int_{-\mu }^\mu
 dy\,\rho (y)  \left( \frac{2}{u-y} - \frac{1}{u-y+M}-\frac{1}{u-y-M} \right),
\end{equation}
which coincides with $\tau '(u)$ up to normalization.

The resolvent has a cut across the eigenvalue interval.  The continuous part has to vanish according to the eigenvalue equation, which is equivalent to the condition 
\begin{equation}\label{cont-cond}
 \mathop{\mathrm{Cont}}r(x)=0,\qquad x\in (-\mu ,\mu ).
\end{equation}
Since its continuous part vanishes, the resolvent equals to $\pm$ its discontinuity on the cut. Defining $r(x)\equiv r(x+i0)$ for definiteness, we have
\begin{equation}\label{fin-diff}
 2\rho (x)-\rho (x-M)-\rho (x+M)=2r(x).
\end{equation}
This is as a second order finite-difference equation for $\rho (x)$. 

Although the equation is the same as (\ref{first-findiff}), the actual problem here is different. Before, the function $r(u)$ was given, being extracted from the supergravity probe analysis. Now $r(u)$ has to be self-consistently determined. We will do it by solving the difference equation for the density, substituting the solution back into the saddle-point equation and reformulating the latter as an integral equation for the resolvent. 

In the limit $M\rightarrow 0$, the difference equation becomes differential:
\begin{equation}
 -M^2\rho ''=2r \qquad (M\rightarrow 0).
\end{equation}
 The solution of the latter is given by the Green's function with Dirichlet boundary conditions at $x=\pm\mu $:
\begin{equation}\label{Green's}
 \rho (x)\stackrel{M\rightarrow 0}{=}\frac{1}{\mu M^2}\int_{-\mu }^{+\mu }dy\,
 \left[
 \theta (x-y)\left(\mu -x\right)\left(\mu +y\right)
 +
 \theta (y-x)\left(\mu +x\right)\left(\mu -y\right)
 \right]r(y).
\end{equation}
When applied to (\ref{sugraprime}), this formula reproduces  (\ref{Wigner1}).

\begin{figure}[t]
\begin{center}
 \centerline{\includegraphics[width=12cm]{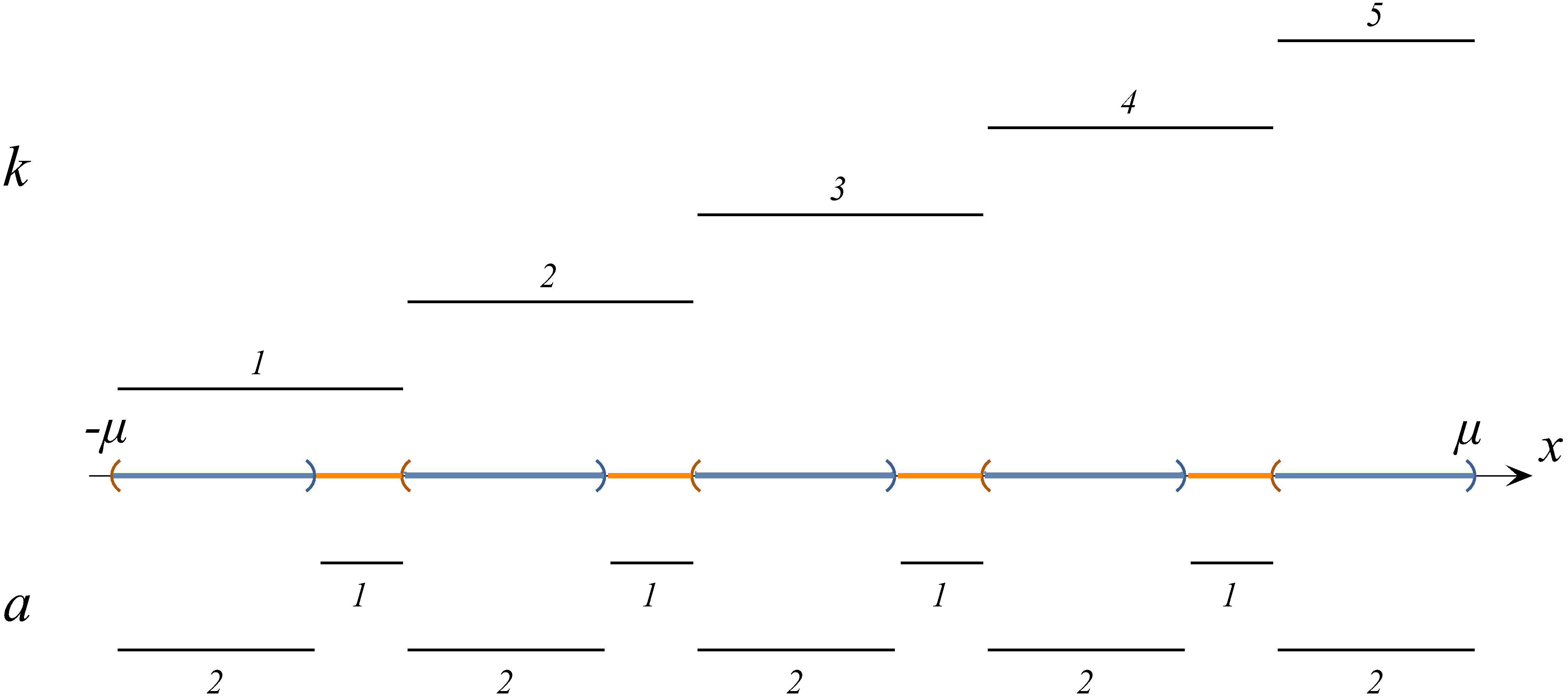}}
\caption{\label{k-a}\small Integers $k$ and $a$ that characterize the Green's function of the difference equation. The interval $(-\mu ,\mu )$ splits in two types of subintervals with $a=2$ and $a=1$.}
\end{center}
\end{figure}

To write down the Green's function of the finite-difference operator, we introduce integer variables $k$ and $a$, illustrated in fig.~\ref{k-a}:
\begin{equation}\label{ka-def}
 k=\left[\frac{\mu +x}{M}\right]+1,\qquad a=\left[\frac{\mu -x}{M}\right]-n+k.
\end{equation}
Then
\begin{equation}\label{disc-Greens}
 \rho (x)=\frac{2k}{n+a}\sum_{l=0}^{n+a-k-l}(n+a-k-l)r(x+lM)+\frac{2(n+a-k)}{n+a}
 \sum_{l=1}^{k-1}(k-l)r(x-lM).
\end{equation}
Replacing sums by integrals we get back to (\ref{Green's}) in the continuum limit.

Exactly the same formula was introduced in \cite{Zarembo:2014ooa} on phenomenological basis, just as an ansatz motivated by the structure of the saddle-point equation (\ref{intnormal1}). We see that this representation naturally arises when eigenvalue distribution is  characterized by the resolvent, thus making direct contact to supergravity.

 The saddle-point equation (\ref{intnormal1}), the finite-difference equation (\ref{fin-diff}) and its solution (\ref{disc-Greens}) are  invariant under simultaneous rescaling $r\rightarrow Cr$, $\rho \rightarrow C\rho $ with constant $C$. We can use this freedom to normalize the resolvent such that near the boundaries of the interval it behaves as
\begin{equation}\label{n-sing}
 r(x)\simeq \frac{1}{\sqrt{M\left(\mu \pm x\right)}}\qquad \left(x\rightarrow \mp\mu \right).
\end{equation}
Fixing the edge behavior is in general inconsistent with canonical normalization of the density. But for our purposes edge normalization is more convenient and we prefer to abandon (\ref{unitnormdens}) in favor of this new condition. The resulting averages, denoted by $\av{\ldots }$, will be wrongly normalized and will have to be divided by a common factor to get correct expectation values:
\begin{equation}\label{physav}
 \left\langle \mathcal{F}(x)\right\rangle=\frac{\av{\mathcal{F}(x)}}{\av{1}}\,.
\end{equation}

\begin{figure}[t]
\begin{center}
 \subfigure[]{
   \includegraphics[width=6cm] {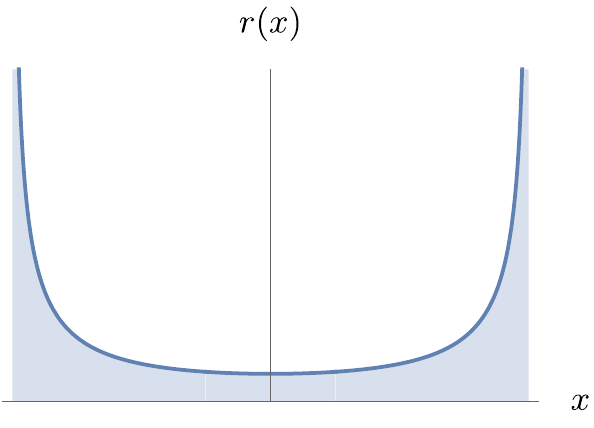}
   \label{fig2:subfig1}
 }
 \subfigure[]{
   \includegraphics[width=6cm] {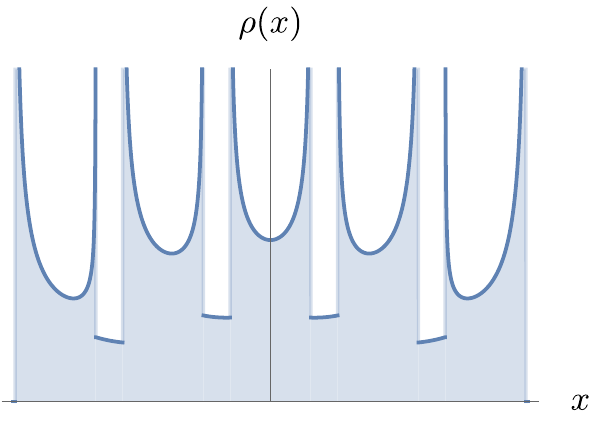}
   \label{fig2:subfig2}
 }
\caption{\label{r-rho}\small From the resolvent (a) to the eigenvalue density (b).}
\end{center}
\end{figure}

While the resolvent is a continuous function,  the discrete map (\ref{disc-Greens}) induces singularities at integer multiples of $M$ away from the endpoints, resulting in a cuspy profile in the middle of the eigenvalue distribution. The detailed structure of cusps is quite intricate. The map from $r$ to $\rho $ splits the interval $(-\mu ,\mu )$ into $2n+1$ subintervals according to the value of $a$, equal to either $1$ or to $2$. The density is a smooth function on the  $a=1$ intervals (``b-intervals", in terminology of \cite{Zarembo:2014ooa,Anderson:2014hxa}) and has inverse square root singularities at the endpoints of the ``a-intervals", those with $a=2$. The overall structure of the map $r(x)\rightarrow \rho (x)$ is illustrated in fig.~\ref{r-rho}.

\subsection{Average formulas}

The next step is to express average quantities in terms of the resolvent.
In computing averages it is convenient to split the integral over $x$ into summation over $k$ defined in (\ref{ka-def}) and integration over $\xi\in (0,1) $ defined as
\begin{equation}
 \xi =\left\{\frac{\mu +x}{M}\right\}.
\end{equation}
Then
\begin{eqnarray}
 \av{\mathcal{F}(x)}&=&\int_{0}^{1}d\xi \,\,\frac{2M}{n+a}\sum_{m,l=1}^{n+a-1}
 r(-\mu +(m-1+\xi )M)\left[
 \theta (l-m)m(n+a-l)
 \right.
\nonumber \\
&&\left.
 +\theta (m-l)l(n+a-m)
 \right]
 \mathcal{F}(-\mu +(l-1+\xi )M),
 \vphantom{\int_{0}^{1}\frac{2M}{n+a}\sum_{m,l=1}^{n+a-1}}
\end{eqnarray}
where, in these variables,
\begin{equation}
 a=1+\theta (\Delta -\xi ).
\end{equation}

The formula looks rather ugly and uninformative but, recalling that the expression in square brackets is the Green's function of the discrete Laplacian, we may anticipate simplifications to occur for functions that are total second derivatives:
\begin{equation}
\hat \Delta \mathcal{F}(x)\equiv 2\mathcal{F}(x)-\mathcal{F}(x+M)-\mathcal{F}(x-M).
\end{equation}
It is indeed easy to show by direct computation that for such functions
summation over $l$ localizes to $l=m$, up to boundary terms at $l=0$ and $l=n+a$:
\begin{eqnarray}
 \av{\hat \Delta \mathcal{F}(x)}&=&\int_{0}^{1}d\xi \,\,\frac{2M}{n+a}\sum_{m=1}^{n+a-1}
 r(-\mu +(m-1+\xi )M)
 \nonumber \\ &&
 \vphantom{\int_{0}^{1}\frac{2M}{n+a}\sum_{m,l=1}^{n+a-1}}
 \times 
 \left[(n+a)\mathcal{F}(-\mu +(m-1+\xi )M)
  \right.
\nonumber \\
&&\left.
 -(n+a-m)\mathcal{F}(-\mu +(\xi -1)M)
  \right.
\nonumber \\
&&\left.
 -m\mathcal{F}(-\mu +(n+a+\xi -1)M)
 \right].
 \vphantom{\int_{0}^{1}\frac{2M}{n+a}\sum_{m,l=1}^{n+a-1}}
\end{eqnarray}
The second line can be written as an integral of a smooth function over the whole interval from $-\mu $ to $\mu $. Changing summation variable in the last term from $m$ to $n+a-m$, and integration variable from $\xi $ to $2-a+\Delta -\xi $ brings the average formula to a very neat form
\begin{equation}\label{averag}
  \av{\hat \Delta \mathcal{F}(x)}=2\int_{-\mu }^{\mu }dx\,r(x)\mathcal{F}(x)
  -\int_{0}^{1}d\xi \,h(\xi)\left(
  \mathcal{F}(-\mu -\xi M)+\mathcal{F}(\mu +\xi M)
  \right),
\end{equation}
where in the last step we also changed $\xi $ to $1-\xi $, and we have introduced a function:
\begin{equation}\label{h-def}
 h(\xi )=\left.\frac{2M}{n+a}\sum_{m=1}^{n+a-1}
 (n+a-m)r(-\mu +(m-\xi )M)\right|_{a=1+\theta (\Delta -1+\xi )}.
\end{equation}

The average formula (\ref{averag}) has a simple meaning.
The first term is the continuum limit that arises upon substituting (\ref{Green's}) for the density, approximating the finite-difference Laplacian by the second derivative and dropping the boundary terms. The last term and the function $h(\xi )$ encode boundary contributions and are the sole remnants of the discreteness inherent to the map (\ref{disc-Greens}).

\begin{figure}[t]
\begin{center}
 \centerline{\includegraphics[width=8cm]{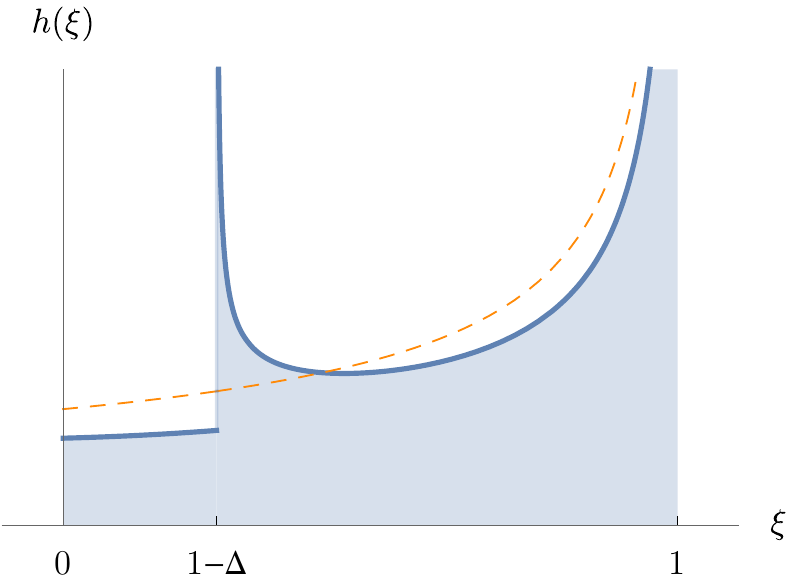}}
\caption{\label{h4}\small The auxiliary function $h(\xi )$ calculated numerically for $n=4$, $\Delta =3/4$. The strong-coupling ($n\rightarrow \infty $) asymptotics $h_\infty (\xi )=2/\sqrt{1-\xi }$ is displayed as a dashed line for comparison.}
\end{center}
\end{figure}

The $h(\xi )$ function will play a prominent r\^ole in the formalism that we are developing and for future use we list here its salient features.
The function is defined on the interval $(0,1)$ and has inverse square root singularities at $\xi =1$ and $\xi =1-\Delta $:
\begin{equation}\label{h-sing}
 h(\xi )\stackrel{\xi \rightarrow 1^-}{\simeq }\frac{n+1}{n+2}\,\,\frac{2}{\sqrt{1-\xi }}\,\qquad 
 h(\xi )\stackrel{\xi \rightarrow (1-\Delta )^+}{\simeq }\frac{1}{n+2}\,\,
 \frac{2}{\sqrt{\xi -1+\Delta }}\,,
\end{equation}
as follows from (\ref{n-sing}). Notice that  when $n\rightarrow \infty $, which corresponds to the strong coupling limit, the singularity at the endpoint is parametrically stronger than the midpoint one.
The overall shape of the function $h(\xi )$ is illustrated in fig.~\ref{h4}.

\subsection{Regular form of integral equation}

The average formula (\ref{averag}) applied to the function $\mathcal{F}(y)=1/(x-y)$ transforms the original saddle-point equation (\ref{intnormal1}) into an equation for the resolvent:
\begin{equation}
 \strokedint_{-\mu }^\mu
 \frac{dy\,r(y)}{x-y}=\int_{0}^{1}d\xi \,h(\xi )\,\frac{x}{x^2-\left(\mu +\xi M\right)^2}\,.
\end{equation}
This equation, first derived in \cite{Zarembo:2014ooa}, is much better behaved than the original integral equation for the density. Now the integral operator has the standard Hilbert kernel and the source term is regular on the whole interval from $-\mu $ to $\mu $. General theorems about singular integral equations \cite{Gakhov} guarantee existence and uniqueness of the solution with boundary conditions (\ref{n-sing}), which in addition is well-behaved in the interior of the eigenvalue interval. The well-known inversion formula for the Hilbert kernel \cite{Gakhov}  yields \cite{Zarembo:2014ooa}:
\begin{equation}\label{resolve}
 r(x)=\frac{1}{\pi \sqrt{\mu ^2-x^2}}
 \int_{0}^{1}d\xi \,h(\xi )\,\frac{\left(\mu +\xi M\right)\sqrt{\xi M \left(2\mu +\xi M \right)}}{\left(\mu +\xi M\right)^2-x^2}\,.
\end{equation}
Having the resolvent, we can formulate the problem entirely in terms of the auxiliary function $h(\xi )$.

\begin{figure}[t]
\begin{center}
 \centerline{\includegraphics[width=8cm]{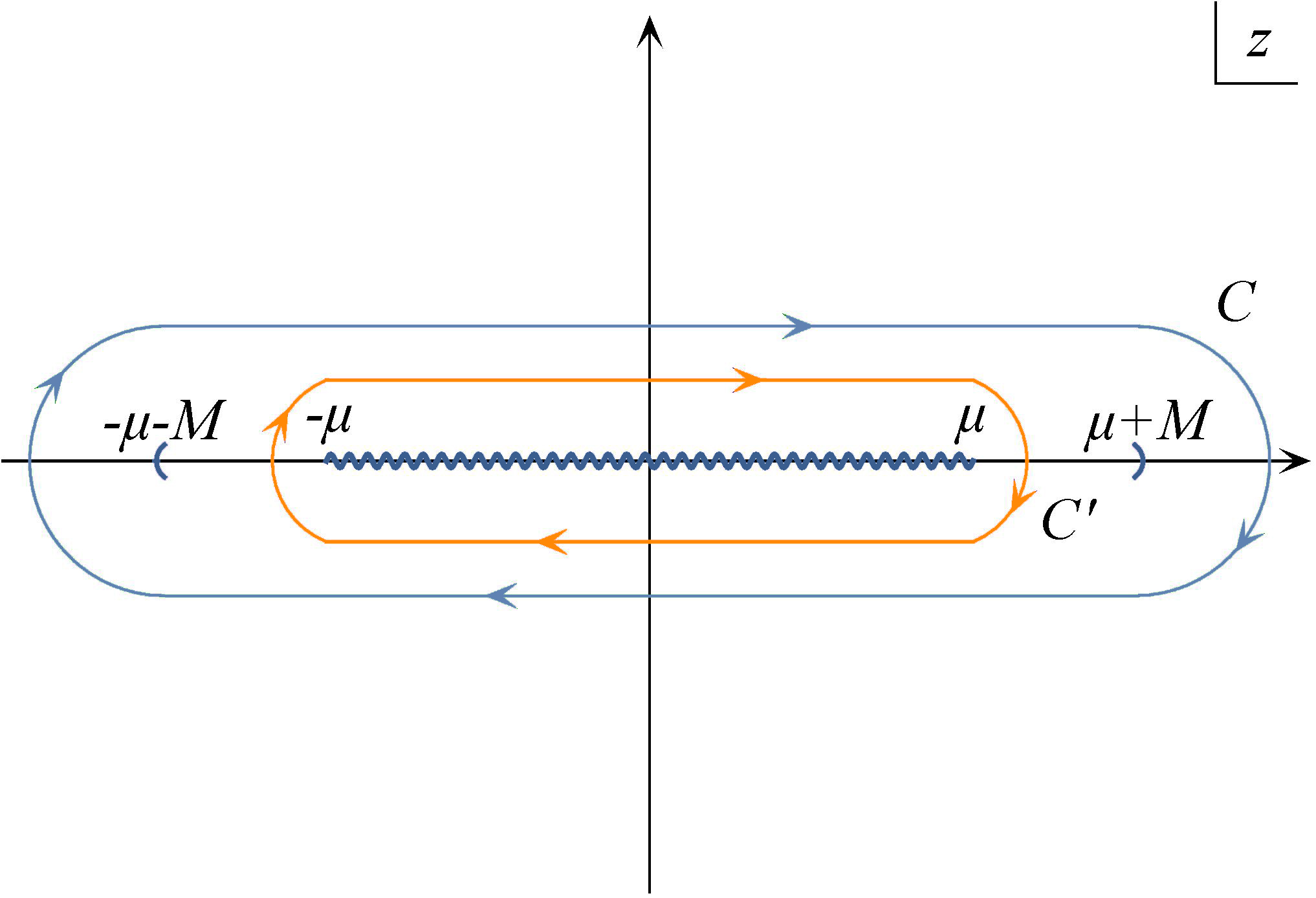}}
\caption{\label{cntrs}\small Contours of integration.}
\end{center}
\end{figure}

We start with the average formula.
Substitution of the integral representation for $r(x)$ into (\ref{averag}) gives
\begin{eqnarray}
 \av{\hat \Delta \mathcal{F}(x)}&=&\int_{0}^{1}d\xi \,h(\xi )\left[
 \left(\mu +\xi M\right)\sqrt{\xi M\left(2\mu +\xi M\right)}
 \vphantom{\int_{-\mu }^{\mu }dx\,\,\frac{\mathcal{F}(x)}{\sqrt{\mu ^2-x^2}}\,\,
 \frac{1}{\left(\mu +\xi M\right)^2-x^2}}
\right. 
\nonumber \\ 
&&\left. \times 
 \,\frac{2}{\pi }
 \int_{-\mu }^{\mu }dx\,\,\frac{\mathcal{F}(x)}{\sqrt{\mu ^2-x^2}}\,\,
 \frac{1}{\left(\mu +\xi M\right)^2-x^2}
 \right.
\nonumber \\
&&\left.
\vphantom{\int_{-\mu }^{\mu }dx\,\,\frac{\mathcal{F}(x)}{\sqrt{\mu ^2-x^2}}\,\,
 \frac{1}{\left(\mu +\xi M\right)^2-x^2}}
 -\mathcal{F}(-\mu -\xi M)-\mathcal{F}(\mu +\xi M)
 \right].
\end{eqnarray}
This result can be further simplified. The integral over $x$ is equivalent to a contour integral, with the contour of integration encircling the cut counterclockwise. Enlarging the contour as shown in fig.~\ref{cntrs}, from $C'$ to $C$, picks  the poles at $z=\pm(\mu +M)$. The residues conspire to cancel the last two terms, leaving a nice compact form of the average formula:
\begin{equation}\label{av1}
 \av{\mathcal{F}(x)}=\frac{1}{M^2}\int_{0}^{1}d\xi \,h(\xi )\widehat{\mathcal{F}}(\xi )
 \left(\mu +\xi M\right)\sqrt{\xi M\left(2\mu +\xi M\right)}\,.
\end{equation}
The hat-operator is defined as
\begin{equation}\label{hat-def}
 \widehat{\mathcal{F}}(\xi )=M^2\oint_C\frac{dz}{\pi i}\,\,
 \frac{\hat \Delta ^{-1}\mathcal{F}(z)}{\sqrt{z^2-\mu ^2}}\,\,
 \frac{1}{z^2-\left(\mu +\xi M\right)^2}\,.
\end{equation}
The contour of integration should leave outside all singularities of $\hat \Delta ^{-1}\mathcal{F}(z)$.

This form of the average formula is ideally suited for the strong-coupling expansion. Inverting the discrete Laplacian is easy for simple functions, and upon computing the contour integral defining $\widehat{\mathcal{F}}(\xi )$ the average reduces to a simple integral with the weight determined by a so far unknown  $h(\xi )$.

Let us illustrate this strategy on some simple examples:
\begin{eqnarray}\label{1-hat}
 &&\hat \Delta ^{-1}\cdot 1=-\frac{x^2}{2M^2}\qquad\Rightarrow\qquad  
 \widehat{1}=1
 \\
 &&\hat \Delta ^{-1}\cdot x^2=\frac{M^2x^2-x^4}{12M^2}\qquad\Rightarrow\qquad 
 \widehat{x^2}=\frac{1}{4}\,\mu ^2+\frac{1}{3}\,M\mu \xi -\frac{1}{6}\,M^2\left(1-\xi \right).
 \label{x2-hat}
 \end{eqnarray}
 For
\begin{equation}
 F(x)\equiv \ln\frac{|M^2-x^2|}{x^2}
\end{equation}
we get:
\begin{equation}\label{F-hat}
  \hat \Delta ^{-1}F(x)=-\ln|x| \qquad\Rightarrow\qquad
 \widehat{F}(\xi )=\frac{2M^2\mathop{\mathrm{arccosh}}\left(1+\frac{\xi M}{\mu }\right)}{\left(\mu +\xi M\right)\sqrt{\xi M\left(2\mu +\xi M\right)}}\,.
\end{equation}

As an application, we can reformulate the normalization condition
(\ref{formintegrated}) in terms of the auxiliary function $h(\xi )$. Since (\ref{formintegrated}) is equivalent to
\begin{equation}\label{lambda-average}
 \frac{8\pi ^2}{\lambda }=\frac{\av{F(x)}}{\av{1}}\,,
\end{equation}
we find:
\begin{equation}\label{forlambdainh-xi}
 \frac{4\pi ^2}{\lambda M^2}\int_{0}^{1}d\xi \,h(\xi )
 \left(\mu +\xi M\right)\sqrt{\xi M\left(2\mu +\xi M\right)}
 =\int_{0}^{1}d\xi \,h(\xi )\mathop{\mathrm{arccosh}}\left(1+\frac{\xi M}{\mu }\right).
\end{equation}
Once $h(\xi )$ is known, this equation would determine $\lambda $ as a function of $\mu /M$.

Another quantity of interest is the vacuum susceptibility:
\begin{equation}\label{vac-s}
 \chi =\frac{16\pi ^{2}}{\lambda M^{2}}\,\left\langle x^2\right\rangle=\frac{16\pi ^{2}}{\lambda M^{2}}\,\,\frac{\av{x^2}}{\av{1}}\,,
\end{equation}
related to the derivative of the partition function with respect to the coupling: $\chi \propto \partial \ln Z/\partial \ln \lambda $. The susceptibility can be computed from (\ref{1-hat}), (\ref{x2-hat}).

All averages can thus be expressed through a single function $h(\xi )$, obtained by applying a linear finite-difference operator (\ref{h-def}) to the resolvent. The resolvent, in its turn, admits an integral representation (\ref{resolve}) in terms of $h(\xi )$. Equations (\ref{resolve}) and (\ref{h-def}) can be solved numerically by iteration. These equations determine $r$ and $h$ up to a common normalization factor, which can be fixed by the asymptotic conditions (\ref{n-sing}) or (\ref{h-sing}), but in practice normalization is not really important because in physical averages an overall factor cancels out, and one can use any convenient normalization whatsoever.

It is possible to eliminate the resolvent from the integral equation and to write down a closed equation for the function $h(\xi )$ only. Applying the difference operator from (\ref{h-def}) to both sides of (\ref{resolve}), we get a linear integral equation
\begin{equation}\label{intforh}
 h(\eta )=\int_{0}^{1}\frac{d\xi }{\pi }\,\,G(\eta ,\xi )h(\xi ),
\end{equation}
where the kernel is given by
\begin{eqnarray}\label{kernelG}
 G(\eta ,\xi )&=&\left.
 \frac{2}{n+a}\sum_{m=1}^{n+a-1}\frac{n+a-m}{m+\xi -\eta }\,\,
 \frac{\mu +\xi M}{2\mu -\left(m-\xi -\eta \right)M}
 \right.
 \vphantom{\sqrt{\frac{\xi }{m-\eta }\,\,
 \frac{2\mu +\xi M}{2\mu -\left(m-\eta \right)M}}}
\nonumber \\
&&\left.\times 
 \sqrt{\frac{\xi }{m-\eta }\,\,
 \frac{2\mu +\xi M}{2\mu -\left(m-\eta \right)M}}\,
 \right|_{a=1+\theta (\Delta -1+\eta )}.
\end{eqnarray}

The formulas (\ref{av1}), (\ref{hat-def}), (\ref{intforh}) and (\ref{kernelG}) completely characterize the eigenvalue distribution, without the need to compute the density nor the resolvent. An arbitrary average can be computed by first solving the integral equation for $h(\xi )$ and then evaluating the integral in (\ref{av1}).  The resolvent and the density can be reconstructed from (\ref{resolve}) and (\ref{disc-Greens}) if necessary, but they are not needed for evaluating expectation values. The system of equations for the eigenvalue distribution written in this form is ideally suited for the strong-coupling expansion.

\section{Strong-coupling expansion}

As anticipated from (\ref{mulambda}) (we will rederive this formula shortly within the formalism described above), $\mu $ grows with $\lambda $ and the strong-coupling regime corresponds to $\mu \gg M$. The relationship between the 't~Hooft coupling and the dimensionless ratio $\mu /M$ follows from  (\ref{forlambdainh-xi}). It  takes zero effort to solve this equation at strong coupling. Expanding  in $M/\mu $ and taking into account that
$$
\mathop{\mathrm{arccosh}}\left(1+\frac{\xi M}{\mu }\right)= \sqrt{\frac{2\xi M}{\mu }}+\mathcal{O}\left(\left(\frac{M}{\mu }\right)^{\frac{3}{2}}\right)\,,
$$
  we get
\begin{equation}
 \frac{4\pi ^2}{\lambda M^2}\,\sqrt{2M\mu ^3}=\sqrt{\frac{2M}{\mu }}\,
 \qquad \left(\lambda \rightarrow \infty \right),
\end{equation}
which confirms (\ref{mulambda}). All the dependence on the function $h(\xi )$ drops out! 

We can find the large-$\lambda $ asymptotics of the vacuum susceptibility, likewise without any detailed knowledge of the function $h(\xi )$.  Keeping only the leading terms in (\ref{x2-hat}), we get:
\begin{equation}
 \chi =\frac{16\pi ^{2}}{\lambda M^{2}}\,\,\frac{\mu ^2}{4}=1\qquad \left(\lambda \rightarrow \infty \right).
\end{equation}

Inspecting the average formulas (\ref{av1}), (\ref{hat-def}) and the kernel (\ref{kernelG}) of the integral equation (\ref{intforh}) one can conclude that
\begin{itemize}
 \item Any expectation value has a regular expansion in $M/\mu $ whose coefficients are expressed through the moments\footnote{Our definition differs from the one in \cite{Zarembo:2014ooa} by an unimportant normalization factor.} of the function $h(\xi )$:
\begin{equation}
 \mathbbm{K}_n=\int_{0}^{1}\frac{d\xi }{\pi }\,h(\xi )\xi ^{n+\frac{1}{2}}.
\end{equation}
 \item  According to (\ref{mulambda}), this expansion translates into a regular expansion in $1/\sqrt{\lambda }$. This structure of the strong-coupling expansion is in accord with expectations from holography, since in the dual description $2\pi/\sqrt{\lambda } $ is identified with the coupling constant of the string sigma-model.
 \item However, the function $h(\xi )$ and consequently its moments depend on $\Delta $ defined in (\ref{n-Delta}), which is clear from fig.~\ref{h4} or from the defining integral equation for $h(\xi )$.
 \item  At strong coupling,
\begin{equation}
 \Delta =\left\{\frac{\sqrt{\lambda }}{\pi }\right\} \qquad \left(\lambda \rightarrow \infty \right).
\end{equation}
 Due to the dependence of $\mathbbm{K}_n$ on $\Delta $, expectation values are not really analytic in $\lambda $, and undergo quantum phase transitions each time $\sqrt{\lambda }$ passes an integer multiple of $\pi $:
\begin{equation}\label{cr-lambda}
 \lambda _c^{(n)}\simeq \pi ^2n^2.
\end{equation}
 This an approximate formula, valid at asymptotically large $\lambda $. An exact equation for the critical points is $2\mu (\lambda _c^{(n)})=nM$ \cite{Russo:2013qaa}.
\end{itemize}

The first strong-coupling correction to $\mu (\lambda ) $ was calculated in \cite{Chen:2014vka,Zarembo:2014ooa} and does not show any non-analyticity in $\lambda $, it was later confirmed by a direct one-loop calculation in the string sigma-model \cite{Chen-Lin:2017pay}. We will push the strong-coupling expansion one order further. Our goal here is two-fold: to  test the formalism developed in the previous section, and  to see if the phase transitions are indeed visible within the strong-coupling expansion.

Expanding (\ref{av1}),  (\ref{1-hat}) -- (\ref{F-hat}) to the third order in $M/\mu $ we get:
\begin{eqnarray}
 \av{1}&=&\pi \sqrt{\frac{2\mu ^3}{M^3}}
 \left(\mathbbm{K}_0+\frac{5\mathbbm{K}_1}{4}\,\,\frac{M}{\mu }+\frac{7\mathbbm{K}_2}{32}\,\,\frac{M^2}{\mu ^2}+\ldots \right)
\nonumber \\
 \av{x^2}&=&\pi \sqrt{\frac{2\mu ^3}{M^3}}\,\frac{\mu ^2}{4}
 \left[
 \mathbbm{K}_0+\frac{31\mathbbm{K}_1}{48}\,\,\frac{M}{\mu }+
 \left(\frac{245\mathbbm{K}_2}{384}-\frac{\mathbbm{K}_0}{6}\right)\frac{M^2}{\mu ^2}+\ldots 
 \right]
\nonumber \\
 \av{F(x)}&=&\pi \sqrt{\frac{2\mu ^3}{M^3}}\,\frac{2M^2}{\mu ^2}
 \left(
 \mathbbm{K}_0-\frac{\mathbbm{K}_1}{12}\,\,\frac{M}{\mu }+
 \frac{3\mathbbm{K}_2}{160}\,\,\frac{M^2}{\mu ^2}+\ldots 
 \right).
\end{eqnarray}
It is convenient, at this point, to introduce the reduced moments, which do not depend on normalization of $h(\xi )$:
\begin{equation}
 \mathbbm{k}_n=\frac{\mathbbm{K}_n}{\mathbbm{K}_0}\,.
\end{equation}
In terms of those, (\ref{lambda-average}) becomes
\begin{equation}\label{lambda-expanded3}
 \frac{4\pi ^2}{\lambda }=\frac{M^2}{\mu ^2}\left[
 1-\frac{4\mathbbm{k}_1}{3}\,\,\frac{M}{\mu }
 +\left(\frac{5\mathbbm{k}_1^2}{3}-\frac{\mathbbm{k}_2}{5}\right)\frac{M^2}{\mu ^2}+\ldots 
 \right].
\end{equation}
Inverting this relationship we obtain:
\begin{equation}\label{mu-NNLO}
 \mu =\frac{\sqrt{\lambda }\,M}{2\pi }
 \left[
 1-\frac{2\mathbbm{k}_1}{3}\,\,\frac{2\pi }{\sqrt{\lambda }}
 +\left(\frac{\mathbbm{k}_1^2}{6}-\frac{\mathbbm{k}_2}{10}\right)\frac{4\pi ^2}{\lambda }+\ldots 
 \right].
\end{equation}

The vacuum susceptibility, to this order in $M/\mu $ is given by
\begin{equation}
 \chi =\frac{4\pi ^{2}\mu ^2}{\lambda M^{2}}
 \left[
 1+\frac{4\mathbbm{k}_1}{3}\,\,\frac{M}{\mu }+
 \left(\frac{7\mathbbm{k}_2}{3}-\frac{5\mathbbm{k}_1^2}{3}-\frac{2}{3}\right)\frac{M}{\mu ^2}+\ldots 
 \right].
\end{equation}
Substituting $\lambda $ using (\ref{lambda-expanded3}), we get:
\begin{equation}
 \chi =1+\left(\frac{32\mathbbm{k}_2}{15}-\frac{16\mathbbm{k}_1^2}{9}-\frac{2}{3}\right)\frac{M^{2}}{\mu ^{2}}+\ldots 
\end{equation}
The term linear  cancels. In terms of the 't~Hooft coupling, 
\begin{equation}\label{chi-strong}
 \chi =
 1
 +\left(\frac{32\mathbbm{k}_2}{15}-\frac{16\mathbbm{k}_1^2}{9}-\frac{2}{3}\right)
 \frac{4\pi ^2}{\lambda }+\ldots 
\end{equation}
Interestingly, the first, NLO correction to the susceptibility vanishes.

In order to compute $\mu $ and $\chi $ to NNLO we need the coefficients $\mathbbm{K}_0$, $\mathbbm{K}_1$, $\mathbbm{K}_2$ to NLO. In fact, the NLO correction enters only for the ratio $\mathbbm{k}_1=\mathbbm{K}_1/\mathbbm{K}_0$. To compute the moments we need to know the function $h(\xi )$, determined by the integral equation (\ref{intforh}). We solve this equation at strong coupling by expanding in $M/\mu $.

Taking the limit $M\rightarrow 0$, $n\rightarrow \infty $ in the kernel (\ref{kernelG}) we get:
\begin{equation}\label{G-infinity}
 G_\infty (\eta ,\xi )=\sum_{m=1}^{\infty }\frac{1}{m+\xi -\eta }\,\sqrt{\frac{\xi }{m-\eta }}\,.
\end{equation}
The solution at this order is \cite{Zarembo:2014ooa}
\begin{equation}\label{asymptotic-h}
 h_\infty (\xi )=\frac{2}{\sqrt{1-\xi }}\,.
\end{equation}
Normalization here is chosen to comply with (\ref{h-sing}), but in principle it is arbitrary, not fixed by the equation, nor affecting any average quantities. It will be important to keep in mind this renormalization ambiguity. Application  to $h_\infty $ of the integral operator with the kernel $G_{\infty }$  returns a telescoping sum, that combines back into $h_{\infty }$:
\begin{equation}\label{G*1/sqrt}
 G_\infty *h_\infty (\eta )=\sum_{m=1}^{\infty }\frac{2}{\sqrt{m-\eta }}\left(1-\sqrt{\frac{m-\eta }{m+1-\eta }}\right)=\frac{2}{\sqrt{1-\eta }}=h_{\infty }(\eta )\,,
\end{equation}
which proves that (\ref{asymptotic-h}) solves  (\ref{intforh}) at the leading order in the strong-coupling expansion.

The moments, computed with the leading-order solution, are
\begin{equation}\label{zero-Kn}
 \mathbbm{K}_n =\frac{2 \Gamma \left(n+\frac{3}{2}\right)}{\sqrt{\pi } (n+1)!}=\frac{2^{-2 n-1} (2 n+2)!}{(n+1)!^2}\ .
\end{equation}
In particular,
\begin{equation}\label{zero-Ks}
 \mathbbm{K}_0 = 1,\qquad \mathbbm{K}_1 = \frac{3}{4}\,,
 \qquad \mathbbm{K}_2 = \frac{5}{8}\,.
\end{equation}
From those we can compute the first-order effective string tension\footnote{The part of the expansion contributed by the leading order $h_{\infty}$ is given by
$$
 \frac{4\pi ^2}{\lambda }=\frac{M^2}{\mu^2}\left[1-\frac{M}{\mu }+\frac{13 M^2}{16 \mu ^2} -\frac{315 M^3}{512 \mu ^3}+...\right]
\ \longrightarrow\ \mu = \frac{\sqrt{\lambda }\,M}{2\pi }\left(1-\frac{\pi }{\sqrt{\lambda }}+\frac{\pi^2}{8\lambda}+\frac{5\pi^3}{128\lambda^{\frac32}}+...\right).
$$
$O(1/\lambda)$ and successive terms will be affected by corrections to (\ref{asymptotic-h}).
} \cite{Chen:2014vka,Zarembo:2014ooa}:
\begin{equation}
 \mu =\frac{\sqrt{\lambda }\,M}{2\pi }\left(1-\frac{\pi }{\sqrt{\lambda }}+\mathcal{O}\left(\frac{1}{\lambda }\right)\right),
\end{equation}
as well as the second-order vacuum susceptibility:
\begin{equation}
 \chi =1-\frac{4\pi ^2}{3\lambda }+\mathcal{O}\left(\frac{1}{\lambda^{\frac{3}{2}} }\right).
\end{equation}

To compute the effective string tension to the second order, we need to know the auxiliary function to the first order in $M/\mu $:
\begin{equation}
 h(\xi )=h_\infty (\xi )+\frac{M}{2\mu }\,h_1(\xi )+\ldots 
\end{equation}
Expanding the kernel (\ref{kernelG}) to the same order,
\begin{equation}
 G(\eta ,\xi )=G_\infty (\eta ,\xi )+\frac{M}{2\mu }\,G_1(\eta ,\xi )+\ldots 
\end{equation}
and substituting into the integral equation (\ref{intforh}), we get an integral equation for $h_1$:
\begin{equation}
 h_1=G_\infty *h_1+G_1*h_\infty .
\end{equation}

The first-order corrections to the kernel (\ref{kernelG}) come from two effects: from $\mathcal{O}(M/\mu )$ corrections to the terms in the sum with $m=\mathcal{O}(1)$ and from the terms with $n-m=\mathcal{O}(1)$. The latter contribution can be obtained by changing summation variable $m\rightarrow n+a-m$ and subsequently extending the summation range to infinity. Altogether we get:
\begin{eqnarray}
 G_1(\eta ,\xi )&=&\frac{3}{2}\,\zeta_{\frac{1}{2}}(1-\eta)\sqrt{\xi }
 +\sum_{m=1}^{\infty }\left(
 \frac{m}{m-a+\Delta +\xi +\eta }\,\sqrt{\frac{\xi }{m-a+\Delta +\eta }}
 \right.
\nonumber \\
&&
 \left.\left.
 -\frac{m}{m+\xi -\eta }\,\sqrt{\frac{\xi }{m-\eta }}
 \right)\right|_{a=1+\theta (\Delta -1+\eta )}.
\end{eqnarray}
Applying it to $h_\infty $ we get:
\begin{equation}\label{ineq-convol}
 G_1*h_\infty =-\frac{1}{2}\,\zeta _{\frac{1}{2}}(1-\eta )+2\zeta _{\frac{1}{2}}(\Delta -\theta (\Delta -1+\eta )+\eta  ).
\end{equation}
Notice that this expression has the same structure of singularities as (\ref{h-sing}).
Using the identity
\begin{equation}
\zeta _{\frac{1}{2}}(\Delta -\theta (\Delta -1+\eta )+\eta  ) =\zeta _{\frac{1}{2}}(\Delta +\eta  )+ \frac{  \theta(\Delta+\eta-1)}{\sqrt{\Delta+\eta-1}}
\end{equation}
we get the explicit form of the equation for $h_1$:
\begin{equation}
 h_1(\eta )=\int_{0}^{1}\frac{d\xi }{\pi }\,\,G_\infty (\eta ,\xi )h_1(\xi )-\frac{1}{2}\,\zeta _{\frac{1}{2}}(1-\eta )+
2\zeta _{\frac{1}{2}}(\Delta +\eta  )+ \frac{2\theta (\Delta+\eta-1)}{\sqrt{\Delta+\eta-1}}.
\label{hhequa}
\end{equation}
There is a freedom of  shifts by the asymptotic solution (\ref{asymptotic-h}):
\begin{equation}\label{shift}
 h_1\rightarrow h_1+Ch_\infty 
\end{equation}
with arbitrary constant $C$. This ambiguity reflects arbitrary normalization of the exact solution. It is easy to check that the normalized averages do not depend on the constant $C$ to the requisite order in $M/\mu $, and so the constant $C$ can be chosen arbitrarily. Iterative solution corresponds to one possible choice.

The first-order correction to the normalized moment
\begin{equation}\label{k1-u}
 \mathbbm{k}_1=\frac{3}{4}\left(1+\frac{M}{2\mu }\,u(\Delta )+\ldots \right)
\end{equation}
is expressed through the solution of the integral equation as
\begin{equation}\label{u-defi}
 u(\Delta )=\int_{0}^{1}\frac{d\xi }{\pi }\,\,h_1(\xi )\left(\frac{4\xi }{3}-1\right)\sqrt{\xi }\,.
\end{equation}
The shift symmetry (\ref{shift}) leaves
this expression invariant.
Substituting this result, along with the unperturbed moments (\ref{zero-Ks}), into (\ref{mu-NNLO}) gives the effective string tension at NNLO of the strong-coupling expansion:
\begin{equation}\label{expansion-of-mu}
 \mu =\frac{\sqrt{\lambda }\,M}{2\pi }\left[
 1-\frac{\pi }{\sqrt{\lambda }}+\left(\frac{1}{8}-u(\Delta )\right)\frac{\pi ^2}{\lambda }+\ldots 
 \right].
\end{equation}
It is  important to note  that at this order $\mu $ starts depending on $\Delta $, which is a fractional part of $\sqrt{\lambda }$ in units of $\pi $. 
This implies that the dependence on $\lambda $ is not  analytic, even though $\mu$  can be expanded in regular power series in the inverse coupling.
The non-analytic behavior occurs at $\Delta=0,1$ and will be discussed in the following section.

The combination of moments that appears in the strong-coupling expansion of susceptibility  (\ref{chi-strong}) is expressed through another function
\begin{equation}\label{v-defi}
 v(\Delta )=\int_{0}^{1}\frac{d\xi }{\pi }\,\,h_1(\xi )\left(\frac{16}{5}\,\xi ^2-4\xi +1\right)\sqrt{\xi }\,.
\end{equation}
We find:
\begin{equation}
 \frac{32\mathbbm{k}_2}{15}-\frac{16\mathbbm{k}_1^2}{9}=\frac{1}{3}+\frac{M}{3\mu }\,v(\Delta )+\ldots 
\end{equation}
Because the susceptibility does not receive corrections at $\mathcal{O}(1/\sqrt{\lambda })$, the dependence on $\Delta $ and the ensuing non-analyticity is postponed by one order in the strong-coupling expansion. We have:
\begin{equation}
 \chi =
 1-\frac{4\pi ^{2}}{3\lambda }+\frac{8\pi ^{3}}{3\lambda ^{\frac{3}{2}}}\left(v(\Delta )+C\right)+\ldots ,
\end{equation}
 where $C$ is some numerical constant that we are not going to compute.

\begin{figure}[t]
\begin{center}
 \subfigure[]{
   \includegraphics[width=6cm] {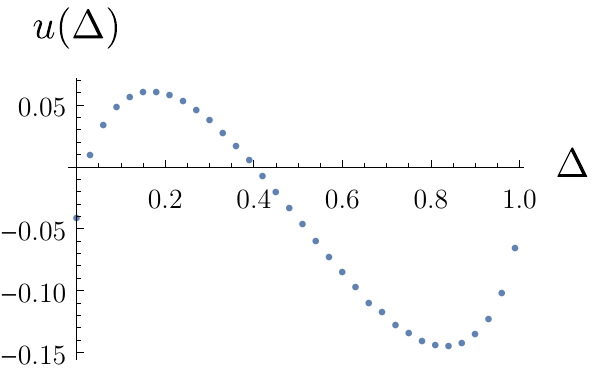}
   \label{figUa}
 }
 \subfigure[]{
   \includegraphics[width=6cm] {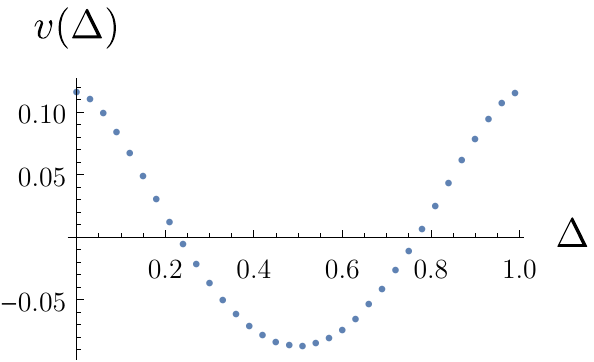}
   \label{figVb}
 }
\caption{\label{graphuv}The functions $u(\Delta)$ and $v(\Delta )$ computed numerically by the method described in appendix~\ref{appendixMoments}.}
\end{center}
\end{figure}

We have calculated the functions $u(\Delta )$ and $v(\Delta )$ numerically (fig.~\ref{graphuv}), solving the integral equation (\ref{hhequa}) by two different methods, either expanding $h_{1}(\xi )$ in the moments
(appendix~\ref{appendixMoments}) or, more directly, by Galerkin method (appendix~\ref{galerkin-app}). The results agree within error bars (see appendix~\ref{galerkin-app}) and demonstrate that both functions depend non-trivially on $\Delta $.

\section{Phase transitions}\label{pt's}

When (\ref{lambda-expanded3}) was inverted to arrive at (\ref{mu-NNLO}) the moments $\mathbbm{k}_{n}$ were assumed to be constant, but in fact they are also functions of 
$\mu $ because of their dependence on $\Delta $. Fortunately, this dependence starts at a rather high order in the strong-coupling expansion and can be taken into account perturbatively. But still, the right-hand side of (\ref{lambda-expanded3}), taken at face value, has different functional dependence on $\mu $ for $\mu =M(n-\epsilon )/2$ and $\mu =M(n+\epsilon )/2$. In the former case, $\Delta =1-\epsilon $, while in the latter, $\Delta =\epsilon $. In practice this means that we should use $u(1-\epsilon )$ in (\ref{k1-u}) slightly below $\mu =Mn/2$ and $u(\epsilon )$ slightly above. In other words, near the critical point functional dependence of the coupling constant on $\mu=M(n+\epsilon ) /2$ is expressed through  $u(\epsilon )$ for $\epsilon >0$ and through $u(1+\epsilon )$ for $\epsilon <0$. Is a function  so defined continuous?  We are going to argue that it is, but that a milder non-analyticity occurs at $\epsilon =0$, leading to critical behavior at $\mu =Mn/2$ or for $\lambda =\lambda _{c}^{(n)}$.

Continuity in $\epsilon $ trivially follows from the integral equation (\ref{ineq-convol}), because there $\Delta $ appears only in combination $\Delta -\theta (\Delta -1+\eta )$, which equals zero for both $\Delta =0$ and $\Delta =1$, independently of $\eta $. The solution  therefore  is the same for $\Delta =0,1$ and so are all the moments of the function  $h_{1}$. The question is what happens slightly below and slightly above the critical point, where
\begin{equation}\label{epsilo}
 \Delta =
\begin{cases}
 \epsilon  & {\rm ~for~ } \epsilon >0
\\
 1+\epsilon  & {\rm ~for~ }\epsilon <0.
\end{cases}
\end{equation}
The analysis of the integral equation (\ref{hhequa}) in the critical region is rather intricate, and we carry out the detailed calculation in the appendix~\ref{near-edge}. Here we present a simplified qualitative argument that illustrates all the salient features.

\subsection{Edge singularity}

We are interested in averages of the form
\begin{equation}\label{<>1}
 \left\langle f(\xi )\right\rangle_{1}=\int_{0}^{1}\frac{d\xi }{\pi }\,\,
 h_{1}(\xi )f(\xi )\sqrt{\xi }\,,
\end{equation}
where $f(\xi )$ is a polynomial. To get a rough idea of what happens for small $\Delta $  or $1-\Delta $, we keep only the source term in the integral equation (\ref{hhequa}):
\begin{equation}
h_1^{(0)}(\eta )=-\frac{1}{2}\,\zeta _{\frac{1}{2}}(1-\eta )+2\zeta _{\frac{1}{2}}(\Delta +\eta  )+ \frac{2\theta (\Delta+\eta-1)}{\sqrt{\Delta+\eta-1}}\,.
\end{equation}
This can be regarded as the zeroth-order approximation in the iterative solution of the integral equation (\ref{hhequa}), albeit there is no small parameter that would justify neglecting the integral term.

It is convenient to use the identity
\begin{equation}
\zeta _{\frac{1}{2}}(\Delta +\eta  ) =\zeta _{\frac{1}{2}}(\Delta +\eta +1 )+ \frac{1}{\sqrt{\Delta+\eta}}\,.
\end{equation}
Taking into account that $\zeta _{\frac{1}{2}}(\Delta +\eta +1 )$ is regular in the full intervals $\eta\in [0,1]$, $\Delta\in [0,1]$, we can write:
\begin{equation}\label{apprh1}
 h_1^{(0)}(\eta )=\frac{2}{\sqrt{\Delta+\eta}}+ \frac{2\theta (\Delta+\eta-1)}{\sqrt{\Delta+\eta-1}}
 +{\rm analytic}.
\end{equation}
For $\Delta \ll 1$, the first term blows up at $\eta \rightarrow 0$ and the second term blows up at $\eta \rightarrow 1$. We can thus set the argument of $f(\xi )$ to zero in the first term and to one in the second term, if we are only interested in the non-analytic behavior at $\Delta \rightarrow 0$:
\begin{eqnarray}\label{d-->0}
 \left\langle f(\xi )\right\rangle_{1}^{(0)}&\eqq& 2f(0)\int_{0}^{1}\frac{d\xi }{\pi }\,\,
 \sqrt{\frac{\xi }{\xi +\Delta }}+2f(1)\int_{1-\Delta }^{1}\frac{d\xi }{\pi}\,\,\frac{1} {\sqrt{\xi -1+\Delta }}
\nonumber \\  &\eqq& \frac{4f(1)}{\pi }\,\sqrt{\Delta }+\frac{f(0)}{\pi }\,\Delta \ln\Delta ,
\end{eqnarray}
where $\eqq$ denotes equality up to an analytic function of $\Delta $. 

When $\Delta \rightarrow 1$, the first term in (\ref{apprh1}) is regular everywhere, while the second term blows up at $\eta \rightarrow 0$. This justifies setting $f(\xi )\simeq f(0)$ and yields:
\begin{equation}\label{d-->1}
 \left\langle f(\xi )\right\rangle_{1}^{(0)}\eqq 2f(0)\int_{1-\Delta }^{1}\frac{d\xi }{\pi}\,\,\sqrt{\frac{\xi } {\xi -1+\Delta }}
 \eqq -\frac{f(0)}{\pi }\,(1-\Delta )\ln(1-\Delta ).
\end{equation}

Next iterations change these results in two ways. The terms containing $\sqrt{\Delta } $ cancel exactly, as we show in appendix~\ref{near-edge}. The logarithmic terms remain, but their coefficients get modified. The exact non-analytic contributions take a neat form in terms of the variable $\epsilon $ introduced in (\ref{epsilo}):
\begin{equation}\label{aveloge}
 \left\langle f(\xi )\right\rangle_{1}\eqq
 \oint\frac{dz }{2\pi ^{2}i}\,\,
 \left(f(z)-f(0)\right)
 \sqrt{\frac{z}{(z-1)^{3}}}\,\,\epsilon \ln|\epsilon |,
\end{equation}
where the contour of integration encircles the interval $(0,1)$ clockwise, leaving singularities of $f(z)$ outside\footnote{The previous simplified calculation neglecting the integral term gives the same formula with $f(z)-f(0)$ replaced by $-f(0)$. Iterations of the integral equation generate the $f(z)$ term.}. For a polynomial this gives:
\begin{equation}
 \left\langle P(\xi )\right\rangle_{1}\eqq
 \frac{1 }{\pi}\,\mathop{\mathrm{res}}_{z=\infty }
 \left(P(z)-P(0)\right)
 \sqrt{\frac{z}{(z-1)^{3}}}\,\,\epsilon \ln|\epsilon |.
\end{equation}
A derivation of these results is given in appendix~\ref{near-edge}.

Applying these findings to $u(\Delta )$ and $v(\Delta )$ from (\ref{u-defi}), (\ref{v-defi}), we get:
\begin{eqnarray}\label{u!}
 u(\Delta )&\eqq& -\frac{2}{\pi }\,\epsilon \ln|\epsilon |
\\
v(\Delta )&\eqq& 0\,\epsilon \ln|\epsilon |.
\label{v!}
\end{eqnarray}
We have solved for the complete $u(\Delta)$, $v(\Delta)$ functions numerically and, in particular, checked that coefficients of the logarithmic terms at $\Delta \rightarrow 0,1$ match with the analytic predictions.

\begin{figure}[t]
\begin{center}
 \subfigure[]{
   \includegraphics[width=6cm] {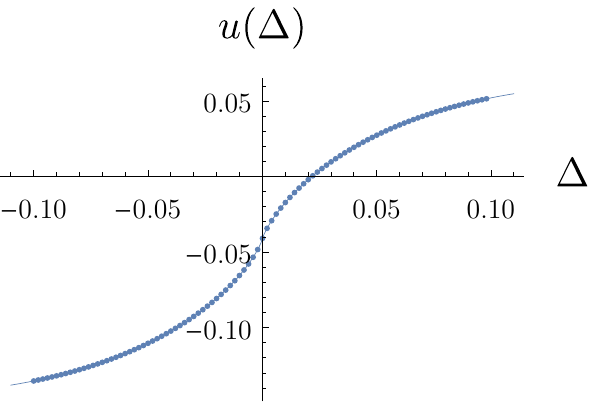}
   \label{figUz}
 }
 \subfigure[]{
   \includegraphics[width=6cm] {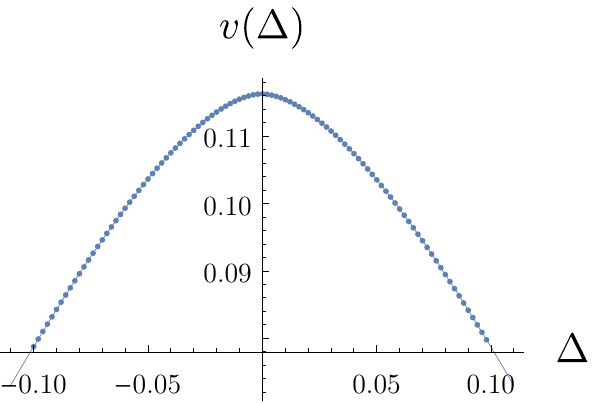}
   \label{figVz}
 }
\caption{$u(\Delta)$ and $v(\Delta )$ computed by the method of appendix \ref{appendixMoments} in the vicinity of $\Delta=0$
and $\Delta=1$. The solid lines correspond to the numerical fits \eqref{uuff}, \eqref{vvff}.}
\end{center}
\end{figure}

The numerical results are shown in figures \ref{figUa} and \ref{figVb}. Figures \ref{figUz}, \ref{figVz} display critical behavior in the vicinity of $\Delta=0$ and $\Delta=1$. The numerical data accurately fits 
the expected $\Delta\log\Delta$  behavior. We find
\bea
u(\Delta )&\approx & -0.04 - 0.55\Delta  - 0.64  \Delta  \log \Delta \ ,\qquad \Delta\to 0^+ 
\label{uuff}\\
\nonumber\\
u(\Delta )&\approx & -0.04 + 0.55(1-\Delta)  + 0.64 (1 - \Delta ) \log(1 - \Delta ) \ ,\qquad \Delta\to 1^- ,
\nonumber
\eea
in good agreement with the analytic result for the coefficient of $\Delta\log\Delta $, $2/\pi\cong 0.637 $.

The function $v(\Delta )$ is less singular. Numerics in this case can be well fit by
\bea
v(\Delta )&\approx & 0.12 + 0.64\Delta^2  +1.91  \Delta^2  \log \Delta \ ,\qquad \Delta\to 0^+ 
\label{vvff}\\
\nonumber\\
v(\Delta )& \approx & 0.12 +0.64 (1-\Delta)^2  +1.91 (1 - \Delta )^2 \log(1 - \Delta ) \ ,\qquad \Delta\to 1^- 
\nonumber
\eea
The absence of the log-linear term is  in agreement with the analytic predictions, and actually follows from the symmetry under $\Delta\rightarrow 1/2-\Delta  $ reflection. The function $v(\Delta )$ is even under reflection, while $u(\Delta )$ is odd upon a shift by a  constant.

The coefficient of the singular term in $v(\Delta )$ is numerically close to $6/\pi \cong 1.910$, and we give the following prediction for the singular part of the function $v(\Delta )$:
\begin{equation}
 v(\Delta )\eqq\frac{6}{\pi }\,\epsilon ^{2}\ln|\epsilon |.
\end{equation}
It would be interesting to derive this result analytically by pushing the perturbative calculation in the appendix~\ref{near-edge} one order further.

\subsection{Critical indices}

The system undergoes phase transitions each time $\Delta $ jumps from $1$ to $0$, which happens at
\begin{equation}
 \mu _{c}^{(n)}=\frac{nM}{2}\,,\qquad \lambda _{c}^{(n)}\simeq \pi^{2}n^{2}.
\end{equation}
We now discuss how non-analyticities that arise at these points affect various  quantities of interest.
Consider first the dynamical scale $\mu $. To find it as a function of the 't~Hooft coupling we need to invert (\ref{lambda-expanded3}). The relevant terms near the critical point are
\begin{equation}\label{approx-lambda-crt}
 \frac{4\pi ^{2}}{\lambda }=\frac{M^{2}}{\mu ^{2}}-\frac{M^{4}}{2\mu ^{4}}\,u(\Delta )+\ldots 
\end{equation}
 where we used (\ref{k1-u}) and omitted unimportant analytic terms. Introducing scaling variables
 $$
\lambda -\lambda _{c}^{(n)}=\delta ,\qquad \mu -\mu _{c}^{(n)}=\frac{M}{2}\,\epsilon ,
$$
expanding (\ref{approx-lambda-crt}) in $\delta $ and $\epsilon $, and using (\ref{u!})  for $u(\Delta )$, we get:
\begin{equation}
 \delta =2\pi ^{2}n\epsilon \left(1-\frac{2}{\pi n}\,\ln|\epsilon |+\ldots \right).
\end{equation}

This formula was derived at large $n\sim \sqrt{\lambda }/\pi$, and  in the course of the derivation we assumed that
the second term in the brackets is a small correction. But near the phase transition this term is logarithmically enhanced and  when $\ln |\epsilon |\sim n$ it is of the same order as the nominally larger leading-order term. Strictly speaking, we cannot infer the behavior of $\mu $ arbitrarily close to the critical point without resumming large logs in the strong-coupling expansion. However, in almost all problems where perturbation theory is logarithmically enhanced,  logarithms exponentiate and log behavior seen in perturbation theory signals power-like scaling with a non-trivial scaling exponent. Taking exponentiation as a plausible assumption,  we can write, to the same degree of accuracy:
\begin{equation}\label{sc-d-e}
 \delta =2\pi ^{2}n\epsilon |\epsilon |^{-\frac{2}{\pi n}+\ldots }
\end{equation}

More generally we can posit power-law scaling of $\mu $ near the critical points:
\begin{equation}
\label{munear}
 \mu -\mu _{c}^{(n)}=\,{\rm const}\,\left(\lambda -\lambda _{c}^{(n)}\right)^{\beta _{n}}.
\end{equation}
This is consistent with the  scaling just derived. Another justification comes from the
exact solution in the weak-coupling phase \cite{Russo:2013qaa} which also exhibits scaling behavior of this form with the critical exponent \cite{Russo:2013kea}:
\begin{equation}
 \beta _{1}=\frac{3}{2}\,.
\end{equation}
From (\ref{sc-d-e}) we infer that at strong coupling:
\begin{equation}
 \beta _{n}=1+\frac{2}{\pi n}+\mathcal{O}\left(\frac{1}{n^{2}}\right),\qquad 
 n\rightarrow \infty \, .
\end{equation}
Interestingly, the strong-coupling approximation is not far off the exact result even for $n=1$.

One can introduce other critical exponents. For example, the vacuum susceptibility is expected to scale as
\begin{equation}
 \chi =\,{\rm const}\,\left(\lambda -\lambda _{c}^{(n)}\right)^{\gamma  _{n}}
 +{\rm analytic}.
\end{equation}
The weak-coupling solution predicts \cite{Russo:2013kea}\footnote{The critical exponent is integer, but it does not mean that susceptibility is analytic. The coefficients in front of the cubic term are different on the two sides of the transition, see  \cite{Russo:2013kea} for more details.}:
\begin{equation}
 \gamma _{1}=3.
\end{equation}
The log-enhancement is lacking in the strong-coupling expansion of the susceptibility, in virtue of (\ref{v!}), which means that at this order $\chi \sim \epsilon $, and the scaling behavior is governed by the same critical exponent as in (\ref{sc-d-e}):
\begin{equation}
 \gamma _{n}=1+\frac{2}{\pi n}+\mathcal{O}\left(\frac{1}{n^{2}}\right),\qquad 
 n\rightarrow \infty. 
\end{equation}
 
Finally, there is a critical exponent associated with the end-point behavior of the density right at the critical point:
\begin{equation}
\label{rhonear}
 \rho _{c}(x)\sim \frac{1}{\left(\mu -x\right)^{\alpha_{n} }} \ ,\qquad \mu\approx \mu_{c}^{(n)}\ .
\end{equation}
This critical exponent has been calculated exactly \cite{Zarembo:2014ooa}:
\begin{equation}
 \alpha_{n} =\frac{1}{2}+\frac{1}{\pi }\,\arcsin\frac{1}{n+1}=\frac{1}{2}+\frac{1}{\pi n}+\mathcal{O}\left(\frac{1}{n^{2}}\right).
\end{equation}
At the first phase transition \cite{Russo:2013kea},
\begin{equation}
 \alpha _{1}=\frac{2}{3}\,.
\end{equation}

While there is no breaking of symmetries at each phase transition, one can define a sequence of ``order parameters" ${\cal O}_n$ of mass dimension one,
that signal the
onset of each  phase transition. They are defined as ${\cal O}_n=1/\ell_n$, where
\be
\ell_n^2 \equiv \left\langle \frac{1}{x^2-(\mu_{c}^{(n)})^2 }\right\rangle
= \int_{-\mu }^\mu dx \rho(x) \frac{1}{x^2-(\mu_{c}^{(n)})^2 }\,.
\ee
The order parameter $\ell_n$ represents a correlation length that diverges at the $n$th-transition point
whenever new resonances occur.
These quantities generalize the similar parameter introduced in 
section 3.5.4 of \cite{Russo:2013kea} for the first ($n=1$) phase transition.
The near-critical behavior can be computed by using (\ref{rhonear}).
The integral has a leading contribution given by
\be
\ell_n^2 \approx 
{\rm const.}\ \left(\mu- \mu_{c}^{(n)} \right)^{-\alpha_{n} }\ .
\ee
Using now \eqref{munear}, we find
\be
{\cal O}_n ={\rm const.}\ \left(\lambda - \lambda_c^{(n)} \right)^{\kappa_n}\ ,\qquad \kappa_n\equiv \frac12 \alpha_n\beta_n\ .
\ee
We see that the  critical exponent $\kappa_n$ is not independent, but derived from $\alpha_n$, i.e.  it is dictated by the behavior
of the eigenvalue density $\rho $ near the endpoint. 
Thus
\be
\kappa_1=\frac12 \ ,\qquad \kappa_n= \frac14+ \frac{1}{n\pi}+\mathcal{O}\left(\frac{1}{n^{2}}\right)\ .
\ee

\section{Conclusions}

The strong-coupling expansion of the effective string tension (\ref{expansion-of-mu}) has an expected structure of perturbative series in the string sigma-model. The first term is reproduced by the classical area law \cite{Buchel:2013id}.
The second term, corresponding to the   one-loop sigma-model correction, was computed by a semi-analytic calculation and it is also in perfect agreement with the NLO of the localization result  \cite{Chen-Lin:2017pay}. The NNLO term then corresponds to the two-loop correction in the sigma-model, and it is there that we expect to see signatures of the phase transitions. 

It is not entirely clear why two loops are sensitive to the phase transitions, while the first two orders are not. Non-analytic dependence on the coupling constant is not that uncommon in quantum field theory and typically arises as a consequence of IR divergences, which entail resummation of perturbative series. Perhaps a similar mechanism is at work here. If so, it must depend on the structure of interactions on the string worldsheet. This is consistent with the observation that at the two lowest orders non-analyticities do not arise, because the classical area law and the one-loop correction due to string fluctuations are not really sensitive to how string modes interact with each other.

Our methods, at least in principle, allow one to develop the strong-coupling expansion in the planar $\mathcal{N}=2^{*}$ theory to any desired order. It would be interesting to carry out higher-order calculations explicitly, in particular to compute the $1/n^{2}$ correction to the critical indices. Even more interesting would be to push perturbative expansion of the string sigma-model \cite{Chen-Lin:2017pay} beyond the one-loop order. Our results predict that quantum phase transitions should be visible on the string side of the holographic duality, once these two-loop corrections are properly taken into account. It is quite remarkable that such a dramatic effect appears to have a perturbative origin in string theory.

Finally, it would be extremely interesting
to understand the phase structure of the $SU(N)$ ${\cal N}=2^*$ theory
for any given $N$.
At finite $N$, the exact analysis of the different phases including instantons has an elegant description in terms of the Seiberg-Witten curve 
\cite{Russo:2014nka,Russo:2015vva,Hollowood:2015oma}.
In the decompactification limit $R\to\infty$ , Pestun's partition function is computed by saddle-points corresponding to
extrema of the action, given by $-R^2{\rm Re} \left(4\pi i {\cal F}\right)$. Thus, the partition function is computed by critical points of the prepotential 
${\cal F}(a_i)$ where the ${\rm Im}(a_{Di})={\rm Im} \frac{\partial {\cal F}}{\partial a_i} $ are required to vanish on the integration domain \cite{Russo:2014nka,Hollowood:2015oma}. 
Such singular points describe massless dyon
singularities.
Phase transitions may occur as the coupling $g_{\rm YM}$ is gradually increased due to the existence of many different competing saddle-points. 
A strong indication that there might be similar phase transitions at any finite $N$, with $N\geq 3$,  was found in \cite{Hollowood:2015oma}, where the saddle-point corresponding to maximal degeneration was found to exist only for $\lambda<\lambda_c$, $\lambda_c\approx 35.42$. Surprisingly, this critical coupling is
the same for any $N$ and corresponds to the first critical point of the large $N$ phase transitions discussed here. An open problem is to identify the singular points that dominate the partition function integral at $\lambda>\lambda_c$ and the total number of different phases for a given $N$. The structure of degenerate points becomes increasingly more difficult  as $N$ is increased, but perhaps for low-rank groups such as $SU(3)$
the different phases can be identified (a discussion of the singularity structure can be found in \cite{Donagi:1995cf}).

\subsection*{Acknowledgments}

We would like to thank B.~Assel, I.~Kostov, J.~Penedones, A.~Sever, J.~Troost and A.~Zhiboedov for discussions. The work of K.Z. and E.W. was supported by the ERC advanced grant No 341222, by the Swedish Research Council (VR) grant
2013-4329, by the grant "Exact Results in Gauge and String Theories" from the Knut and Alice Wallenberg foundation, and by RFBR grant 18-01-00460 A. 
J.G.R. acknowledges financial
support from projects 2017-SGR-929, MINECO grant FPA2016-76005-C.

\appendix

\section{Computation of \texorpdfstring{$u(\Delta )$}{u(Delta)} and \texorpdfstring{$v(\Delta)$}{v(Delta)} }\label{appendixMoments}

Our starting point is $G_\infty(\eta,\xi)$,
\bea
G_\infty(\eta,\xi)
&=& \sum_{k=1}^\infty \frac{1}{k+\xi-\eta} \frac{\sqrt{\xi}}{\sqrt{k-\eta}}
\nonumber\\
&=& \frac{1}{1+\xi-\eta} \frac{\sqrt{\xi}}{\sqrt{1-\eta}}+
\sum_{k=2}^\infty \frac{1}{k+\xi-\eta} \frac{\sqrt{\xi}}{\sqrt{k-\eta}} \, .
\label{ggff}
\eea
We have separated the term $k=1$. The terms with $k\geq 2$  can be expanded
in powers of $\xi $ and the expansion converges for all $\eta\in [0,1]$.  
The remaining sum from $k=2$ to $\infty$ can then be expressed in terms of  Hurwitz $\zeta$ functions.
We obtain
\be
G_\infty(\eta,\xi)= \frac{1}{1+\xi-\eta} \frac{\sqrt{\xi}}{\sqrt{1-\eta}} +\sum_{r=0}^\infty (-1)^r \xi^{r+\frac12}\zeta_{r+\frac32}(2-\eta)
\, .
\ee
Now we consider the equation (\ref{hhequa}) for $h_1(\eta)$.
Using the above expansion of $G_\infty(\eta,\xi)$, we find
\bea
h_1(\eta )&=& \int_0^1 \frac{d\xi}{\pi}  \frac{ h_1(\xi )}{1+\xi-\eta} \frac{\sqrt{\xi}}{\sqrt{1-\eta}}+  \sum_{r=0}^\infty (-1)^r \tilde k_r \zeta_{r+\frac12}(2-\eta)   
\nonumber\\
&-&\frac12 \zeta_{\frac12}(1-\eta)+2\zeta _{\frac{1}{2}}(\Delta +\eta  )+ \frac{2\theta (\Delta+\eta-1)}{\sqrt{\Delta+\eta-1}}
\label{eqas}
\eea
where 
\be
\tilde k_r = \int_0^1 \frac{d\xi}{\pi} h_1(\xi) \xi^{r+\frac12}\ .
\ee
This equation can be converted into a linear algebraic equation for the moments by 
 multiplying by $\eta^{n+1/2}$ and integrating over $\eta $. 
All the integrals  are convergent.
The first integral may be computed by residues. 
The result  is given by the following formula:
\be
   \int_0^1 \frac{d\eta}{\pi} \frac{\eta^{n+1/2}}{1+\xi-\eta} \frac{\sqrt{\xi}}{\sqrt{1-\eta}} = 
  (1+\xi)^{n+1/2}-\sum_{r=0}^n c_r  
  (1+\xi)^{n-r}\ ,
\ee
where
\be
c_r\equiv \frac{(-1)^r\sqrt{\pi}}{r!\Gamma(\frac12-r )} \, .
\ee
Expanding in powers of $\xi $, the integral over $\xi $ can be computed in terms of moments.
We find 
\be
 \int_0^1 \frac{d\xi}{\pi}  h_1(\xi )  \int_0^1 \frac{d\eta}{\pi} \frac{\eta^{n+1/2}}{1+\xi-\eta} \frac{\sqrt{\xi}}{\sqrt{1-\eta}} =\sum_{s=0}^\infty f_{s,n} \tilde k_{s-\frac12}+
 \sum_{s=0}^n  q_{s,n}\tilde k_{s}\ .
\ee
with
\be
q_{s,n} \equiv \frac{2^{-2 n-1} (-1)^s (2 n+1)! \Gamma \left(-s-\frac{1}{2}\right)}{\sqrt{\pi } n! (n-s)!}\ ,\qquad f_{s,n}\equiv \frac{\Gamma(n+\frac32 )}{s! \Gamma(n+\frac32 -s)}\ .
\ee
Note that the first term contains moments
\be
\tilde k_{r-\frac12 } = \int_0^1 \frac{d\xi}{\pi} h_1(\xi) \xi^{r}\ .
\ee
A second  equation for the moments $\tilde k_{r-1/2}$ can be obtained 
by performing the integration $ \int d\eta\ \eta^ n  $ in (\ref{eqas}).
We need the formula:
\bea
&& \int_0^1 \frac{d\xi}{\pi}  h_1(\xi )  \int_0^1 \frac{d\eta}{\pi} \frac{\eta^{n}}{1+\xi-\eta} \frac{\sqrt{\xi}}{\sqrt{1-\eta}} =
\nonumber\\
= && \int_0^1 \frac{d\xi}{\pi}  h_1(\xi ) 
\frac{n!  \sqrt{\xi}}{\sqrt{\pi }(1+\xi)
   \Gamma \left(n+\frac{3}{2}\right)} \, _2F_1\left(1,n+1;n+\frac{3}{2};\frac{1}{1+\xi }\right)
\\
= && \int_0^1 \frac{d\xi}{\pi}  h_1(\xi ) 
\left(\sum_{r=0}^n g_{r,n} \xi^{r}+\sum_{r=0}^\infty g_{r+\frac12,n} \xi^{r+\frac12}\right)
\\
= && 
\sum_{r=0}^n g_{r,n} \tilde k_{r-1/2}+\sum_{r=0}^\infty g_{r+\frac12,n} \tilde k_{r} \, .
\eea
For integer $n$, the hypergeometric function 
 reduces to an arcsin function combined with square roots.

We thus obtain the following linear system of equations for
moments $\{ \tilde k_r,\tilde k_{r+1/2} \}$ 
\bea
\tilde k_n &=& \sum_{r=0}^\infty (-1)^r c_{r,n} \tilde k_r+\sum_{r=0}^n  q_{r,n}\tilde k_{r}+
\sum_{r=0}^\infty f_{r,n} \tilde k_{r-\frac12} -\frac12 \ell_n +2 p_n
\nonumber\\
\label{eqal}
\\
\tilde k_{n-\frac12} &=&  \sum_{r=0}^\infty (-1)^r c_{r,n-\frac12} \tilde k_r
+\sum_{r=0}^n g_{r,n} \tilde k_{r-\frac12 }+\sum_{r=0}^\infty g_{r+\frac12,n} \tilde k_{r}
-\frac12 \ell_{n-\frac12} +2p_{n-\frac12} 
\nonumber
\eea
where 
\be
c_{r,n}= \int_0^1 \frac{d\eta}{\pi} \eta^{n+1/2} \zeta_{r+\frac32 }(2-\eta )\ ,\qquad
\ell_n=\int_0^1 \frac{d\eta}{\pi} \eta^{n+1/2} \zeta_{\frac12 }(1-\eta )
\, .
\ee
\bea
p_n &=&\int_0^{1-\Delta} \frac{d\eta}{\pi} \eta^{n+1/2} \zeta_{\frac12 }(\Delta+\eta )+ 
\int_{1-\Delta}^1 \frac{d\eta}{\pi} \eta^{n+1/2} \zeta_{\frac12 }(\Delta+\eta-1 )
\nonumber\\
&=&\int_0^{1} \frac{d\eta}{\pi} \eta^{n+1/2} \zeta_{\frac12 }(\Delta+\eta+1 )+ 
\int_{0}^1 \frac{d\eta}{\pi} \frac{ \eta^{n+1/2} }{\sqrt{\Delta+\eta}}+
\int_{1-\Delta}^1 \frac{d\eta}{\pi} \frac{ \eta^{n+1/2} }{\sqrt{\Delta+\eta-1}}
\, .
\nonumber
\eea
Equations (\ref{eqal}) can be solved as an  algebraic system of linear equations by
truncating the infinite sum to some maximum value $n_{\rm max}$.
Some of the coefficients involve $\Gamma $ functions and become large for large values of $r,n$.
These coefficients are however multiplied by the moments $\tilde k_r$ or $\tilde k_{r-\frac12 }$, which for large $r$ are very small. It is in fact more convenient to solve the system by iterations, beginning with vanishing values for
$\{ \tilde k_{-1/2},\tilde k_0,....,\tilde k_{N-1/2},\tilde k_{N}\} $, since the direct solution of the linear algebraic equations has to deal with a matrix with huge coefficients due to the above $\Gamma $ functions.
The moments converge very rapidly after a few iterations.

\section{Galerkin Method for \texorpdfstring{$u(\Delta)$}{u(Delta)} and \texorpdfstring{$v(\Delta)$}{v(Delta)}}
\label{galerkin-app}
\newcommand{\fgal}{f}
As a consistency check of the numerical results for $u(\Delta)$ and $v(\Delta)$, we also made a numerical approximation of the integral equation \eqref{hhequa} for $h_1$ by use of the Galerkin method. The equation was thus projected to the finite dimensional function space spanned by the following $6+8$ basis functions $\fgal_i$, piecewisely defined on two subintervals:
\begin{center}
  \begin{tabular}{c | c}
    $\eta \in [0, 1-\Delta]$	&	$\eta \in (1-\Delta, 1]$ \\ \hline \rule{0pt}{1.7ex}
    six first Chebyshev polynomials	& six first Chebyshev polynomials	\\
    &	together with $\frac{1}{\sqrt{1-\eta}}$ and $\frac{1}{\sqrt{\eta - (1-\Delta)}}$.
  \end{tabular}
\end{center}
These basis functions were orthonormalized with respect to the scalar product $\galscal{\cdot, \cdot}$ with weight function 
\begin{equation}
  w(\eta) = \begin{cases}
    \sqrt{1 - \eta^2} 			\;, 	\qquad &\eta \in [0,1-\Delta]	\\
    \big(1-\eta\big)\big(\eta - (1-\Delta) \big) \; , \qquad  & \eta \in (1-\Delta, 1]
  \end{cases}
  .
\end{equation}
The Galerkin method then reduces the integral equation to a linear equation system for the best approximating coefficients in $h_1(\eta) \approx  \sum_i c_i \fgal_i(\eta)$:
\begin{gather}
  c_i =  \sum_j A_{ij} c_j + b_i \;, \qquad {\text{where}}
  \\
  A_{ij} = \int \! \frac{d\xi}{\pi} \; \galscal*{ \fgal_i, G_{\infty}}(\xi) \; \fgal_j(\xi)
  \\
  b_i = \galscal*{\fgal_i\; , \; -\frac{1}{2} \zeta_{\frac{1}{2}}(1-\eta) + 2 \zeta_{\frac{1}{2}}(\Delta+\eta) + \frac{2\theta(\Delta+\eta-1)}{\sqrt{\Delta+\eta-1}}}
  .
\end{gather}

\begin{figure}[t]
  \centering
 \subfigure[]{
   \includegraphics[width=6cm] {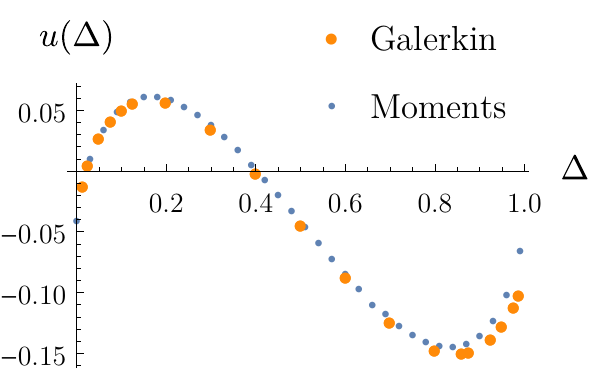}
   \label{fig:galerkinU}
 }
 \subfigure[]{
   \includegraphics[width=6cm] {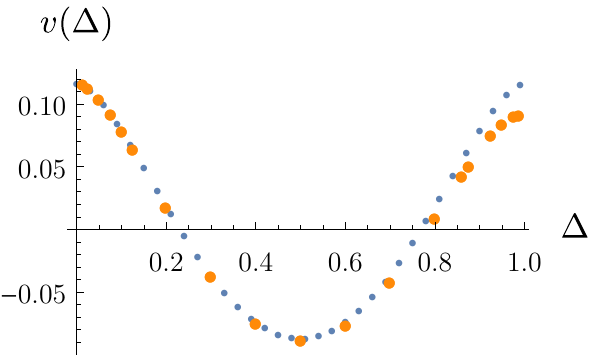}
   \label{fig:galerkinV}
 }
 \caption{\label{fig:galerkin} $u(\Delta)$ and $v(\Delta )$ as computed by the Galerkin method. The moments based method of appendix \ref{appendixMoments} is plotted with smaller dots as comparison. There is agreement within the accuracy except as $\Delta$ approaches 1 where the Galerkin method becomes increasingly inaccurate due to the singular basis functions.}
\end{figure}

The integrals were computed numerically using interpolation of the kernel $G_{\infty}(\eta, \xi)$. The results are shown in figure \ref{fig:galerkin}, together with those of the numerical approximation of the moments presented in the main text.

\section{Living at the edge}
\label{near-edge}

Here we complement qualitative arguments of  section~\ref{pt's} by a more rigorous analysis of the integral equation (\ref{hhequa}). It follows from the discussion in the main text that the non-analytic terms at  $\Delta\rightarrow 0,1$ arise either from parametrically small $\xi $ (the $\ln\Delta $ terms) or  from $\xi $ very close to one (the $\sqrt{\Delta }$ term). The expected singularities were estimated from the source term in the integral equation, disregarding higher-order iterations. This approximation cannot be really justified, and below we determine asymptotics of the exact solution near $\eta =0$ and $\eta =1$ without making any uncontrollable approximations. The end-point singularities of the source  are captured by (\ref{apprh1}), but there is also an integral term in (\ref{hhequa}) which was neglected in the simplified calculation of  section~\ref{pt's}.

If we set $\eta = 0$ in the kernel of the integral equation, nothing dramatic happens. Explicitly, from (\ref{G-infinity}) we get:
\begin{equation}
 G_{\infty }(0,\xi )=\sum_{m=1}^{\infty }\frac{1}{m+\xi }\,\sqrt{\frac{\xi }{m}}\,,
\end{equation}
which is a nice, regular function on the whole interval $(0,1)$. The singularity at $\eta =0$, therefore, is entirely determined by the source. Introducing the variable $\epsilon $ from (\ref{epsilo}) we can recast (\ref{apprh1}) in the form valid for both $\epsilon >0$ ($\Delta\rightarrow 0 $) and $\epsilon <0$ ($\Delta \rightarrow 1$):
\begin{equation}\label{near-0}
 h_{1}(\eta )\stackrel{\eta \rightarrow 0}{\simeq} 2\,\frac{\theta (\eta +\epsilon )}{\sqrt{\eta +\epsilon }}\,. 
\end{equation}

On the contrary, setting $\eta =1$ in (\ref{G-infinity}) yields a function with a singularity at $\xi =0$:
\begin{equation}
 G_{\infty }(\eta ,\xi )\stackrel{\xi \rightarrow 0}{\stackrel{\eta \rightarrow 1}{\simeq }}
 \frac{1}{1+\xi -\eta }\,\sqrt{\frac{\xi }{1-\eta }}\,.
\end{equation}
The integral term in the equation will thus get a singular contribution from $\xi $ very close to zero.
But we already know the behavior of $h_{1}(\xi )$ at the left end of the interval, it is given by (\ref{near-0}).  Adding this integral contribution to the singular part of the source term we get\footnote{We need to impose a cutoff on the  $\xi $-integration, because the approximate formula (\ref{near-0}) is only valid for $\eta \ll 1 $. We assume that $|\epsilon |\ll \ell\ll 1$.}:
\begin{equation}\label{exactly_right}
  h_{1}(\eta )\stackrel{\eta \rightarrow 1}{\simeq} 
  2\,\frac{\theta (\eta -1+\epsilon )}{\sqrt{\eta -1+\epsilon }}
  +2\int_{0}^{\ell}\frac{d\xi }{\pi }\,\,
  \frac{1}{1+\xi -\eta }\,\sqrt{\frac{\xi }{1-\eta }}\,\,
  \frac{\theta (\xi  +\epsilon )}{\sqrt{\xi  +\epsilon }}
  \,.
\end{equation}
The integral evaluates to elementary functions, but for our purposes the integral representation is more convenient.

Equipped with the asymptotic form of the solution at the two extremities of the interval, we can now compute the boundary contribution to the average (\ref{<>1}). The source term at the right boundary gives $\sqrt{\epsilon }$ for $\epsilon >0$, as shown in (\ref{d-->0}). Having the integral term at our disposal we add the two contributions together. Assuming $\epsilon >0$, we get from (\ref{exactly_right}):
\begin{eqnarray}
 \left\langle f(\xi )\right\rangle_{1}^{\xi \rightarrow 1}
 &\eqq& 2 f(1)\left(
 \int_{0}^{\epsilon }\frac{d\nu }{\pi \sqrt{\epsilon -\nu }}
 +\int_{0}^{\ell}\frac{d\xi }{\pi }\,\,\sqrt{\frac{\xi }{\xi +\epsilon }}
 \int_{0}^{\infty }\frac{d\nu }{\pi \sqrt{\nu }}\,\,\frac{1}{\nu +\xi }
 \right)
\nonumber \\
&\eqq&
 \frac{4f(1)}{\pi }\left(\sqrt{\epsilon }+\sqrt{\ell+\epsilon }-\sqrt{\epsilon }\right)
 \eqq 0
 \vphantom{+\int_{0}^{\ell}\frac{d\xi }{\pi }\,\,\sqrt{\frac{\xi }{\xi +\epsilon }}}
 \qquad \left(\epsilon >0\right).
\end{eqnarray}
The square root has completely cancelled! For $\epsilon <0$ the source term is not singular and the integral term does not induce any new singularities either:
\begin{equation}
  \left\langle f(\xi )\right\rangle_{1}^{\xi \rightarrow 1}\eqq
  \frac{2f(1)}{\pi }\int_{|\epsilon |}^{\ell}\frac{d\xi }{\sqrt{\xi -|\epsilon |}}
  \eqq 0 \qquad \left(\epsilon <0\right).
\end{equation}

The singular contribution from the other end was already computed in (\ref{d-->0}), (\ref{d-->1}):
\begin{equation}\label{md-->0}
 \left\langle f(\xi )\right\rangle_{1}^{\xi \rightarrow 0}
 \eqq 
 \frac{f(0)}{\pi }\,\epsilon \ln|\epsilon |.
\end{equation}
The integral term does not contribute, 
but this is not the end of the story,  because iterations of the integral equation induce a small but non-analytic piece in the bulk:
\begin{equation}\label{hhatbulk}
 h_{1}(\eta )\eqq \widehat{h}(\eta )\epsilon \ln|\epsilon |.
\end{equation}

Indeed, the integral equation (\ref{hhequa}) can be written as
\begin{equation}
 h_{1}(\eta )=\left\langle \frac{G_{\infty }(\eta ,\xi )}{\sqrt{\xi }}\right\rangle_{1}+{\rm source}.
\end{equation}
Taking into account that 
\begin{equation}
 \lim_{\xi \rightarrow 0}\frac{G_{\infty }(\eta ,\xi )}{\sqrt{\xi }}
 =\zeta _{\frac{3}{2}}(1-\eta ),
\end{equation}
and applying (\ref{md-->0}), we find that $\widehat{h}$ should satisfy the following integral equation:
\begin{equation}\label{inteq-C}
 \widehat{h}(\eta )=\frac{1}{\pi }\,\zeta _{\frac{3}{2}}(1-\eta )
 +\int_{0}^{1}\frac{d\xi }{\pi }\,\,G_{\infty }(\eta ,\xi )\widehat{h}(\xi ).
\end{equation}

The equation can be solved with the help of the identity
\begin{equation}\label{G*32}
 G_{\infty }*\frac{1}{(1-\xi)^{{\frac{3}{2}}} }=\frac{1}{(1-\eta )^{\frac{3}{2}}}
 -\zeta _{{\frac{3}{2}}}(1-\eta ),
\end{equation}
that can be proved by the same of chain of arguments as (\ref{G*1/sqrt}). The convolution integral here should be understood in the analytic sense as a contour integral around the cut, which removes an apparent divergence at $\xi =1$. The divergence arises because near $\xi =1$ the bulk solution is no longer accurate and has to be replaced with the asymptotic expression (\ref{exactly_right}).

Substituting (\ref{G*32}) into (\ref{inteq-C}) we find:
\begin{equation}
 \widehat{h}(\eta )=\frac{1}{\pi (1-\eta )^{\frac{3}{2}}}\,.
\end{equation}
The total non-analytic piece of the average combines the contribution from the boundary (\ref{md-->0}) with the contribution from the bulk (\ref{hhatbulk}):
\begin{equation}
 \left\langle f(\xi )\right\rangle_{1}
 \eqq
 \left[\frac{f(0)}{\pi }+\int_{0}^{1}\frac{d\xi }{\pi ^{2}}\,\,
 f(\xi )\sqrt{\frac{\xi }{(1-\xi )^{3}}}\right]\epsilon \ln|\epsilon |.
\end{equation}
The nominally divergent integral should  again be understood in the analytic sense.
In the main text we use an equivalent contour-intergal representation.

\bibliographystyle{nb}

\end{document}